\documentclass[11pt]{article}
\usepackage[T1]{fontenc}
\usepackage[margin=1in]{geometry}
\usepackage{amsmath,amssymb}
\PassOptionsToPackage{type1,nofontspec}{newtxtext}
\usepackage{newtxtext,newtxmath}
\ifdefined\pdfglyphtounicode
\pdfglyphtounicode{parenleftbigg}{0028}
\pdfglyphtounicode{parenrightbigg}{0029}
\pdfglyphtounicode{floorleftbigg}{230A}
\pdfglyphtounicode{floorrightbigg}{230B}
\fi
\usepackage{booktabs,siunitx,graphicx,array,longtable}
\usepackage{subcaption}
\usepackage{placeins}
\usepackage{needspace}
\usepackage[round,authoryear]{natbib}
\usepackage{xcolor}
\usepackage{fancyhdr}
\usepackage{microtype}
\usepackage{xurl}
\usepackage[colorlinks=true,linkcolor=blue!45!black,citecolor=blue!45!black,urlcolor=blue!45!black]{hyperref}
\newcommand{\doi}[1]{\url{https://doi.org/#1}}
\newcommand{\IdentifierBreaks}{\def\UrlBreaks{\do\/\do\-\do\_\do\?\do\&\do\=\do\#}\def\UrlBigBreaks{\do\:}}
\AddToHook{cmd/section/before}{\Needspace{6\baselineskip}}
\AddToHook{cmd/subsection/before}{\Needspace{5\baselineskip}}
\AddToHook{cmd/paragraph/before}{\Needspace{3\baselineskip}}
\newcommand{\TableHeadingSpace}{\Needspace{12\baselineskip}}
\newcommand{\ValidationTableSpace}{\Needspace{18\baselineskip}}
\newcommand{\RainfallTableSpace}{\Needspace{16\baselineskip}}
\newcommand{\CovariateTableSpace}{\Needspace{20\baselineskip}}
\newenvironment{WideTablePanel}{\begin{minipage}[t]{0.53\linewidth}\vspace{0pt}}{\end{minipage}}
\newenvironment{NarrowTablePanel}{\begin{minipage}[t]{0.45\linewidth}\vspace{0pt}}{\end{minipage}}
\newcommand{\AlignedPanelCaption}[1]{\begin{minipage}[t][6\baselineskip][t]{\linewidth}#1\end{minipage}\par}

\fancypagestyle{plain}{\fancyhf{}\fancyfoot[C]{\thepage}}
\newcommand{\AcquisitionDays}{5,099}
\newcommand{\AcquisitionEnd}{2026-01-04}
\newcommand{\AcquisitionStart}{2012-01-20}
\newcommand{\AllConfidenceDetections}{3,842,904}
\newcommand{\BlockDegrees}{1}
\newcommand{\BoostRounds}{400}
\newcommand{\BootstrapReplicates}{500}
\newcommand{\CCVersion}{4.0}
\newcommand{\CalendarWeeks}{52}
\newcommand{\CalibHeldPositivePercent}{2.29}
\newcommand{\CalibLabelPercent}{92.16}
\newcommand{\CalibPositivePercent}{3.04}
\newcommand{\CalibRows}{1,894,446}
\newcommand{\CalibYear}{2022}
\newcommand{\CalibratedBaselineContrast}{[\num{\CalibratedBaselineGainLow}, \num{\CalibratedBaselineGainHigh}]}
\newcommand{\CalibratedBaselineGainHigh}{0.2212}
\newcommand{\CalibratedBaselineGainLow}{0.1027}
\newcommand{\CalibratedClimatologyAP}{0.163}
\newcommand{\CalibratedClimatologyBrier}{0.018}
\newcommand{\CalibratedClimatologyECE}{0.0088}
\newcommand{\CalibratedPersistenceAP}{0.143}
\newcommand{\CalibratedPersistenceBrier}{0.017}
\newcommand{\CalibratedPersistenceECE}{0.010}
\newcommand{\ChirpsMajor}{3}
\newcommand{\ChirpsRelease}{2025-01-01}
\newcommand{\ChirpsVersion}{3.0}
\newcommand{\ComparisonBootstrapReplicates}{500}
\newcommand{\ComparisonCNNCalibAP}{0.314}
\newcommand{\ComparisonCNNGain}{-0.0026}
\newcommand{\ComparisonCNNGainHigh}{0.0035}
\newcommand{\ComparisonCNNGainLow}{-0.0070}
\newcommand{\ComparisonCNNSelectAP}{0.350}
\newcommand{\ComparisonCNNTestCalibratedAP}{0.319}
\newcommand{\ComparisonCNNTestRawAP}{0.324}
\newcommand{\ComparisonCalibrationBlocks}{10}
\newcommand{\ComparisonCatBoostCalibAP}{0.319}
\newcommand{\ComparisonCatBoostGain}{0.0031}
\newcommand{\ComparisonCatBoostGainHigh}{0.0096}
\newcommand{\ComparisonCatBoostGainLow}{-0.0019}
\newcommand{\ComparisonCatBoostSelectAP}{0.349}
\newcommand{\ComparisonCatBoostTestCalibratedAP}{0.331}
\newcommand{\ComparisonCatBoostTestRawAP}{0.337}
\newcommand{\ComparisonCatCPUSelectionAP}{0.347}
\newcommand{\ComparisonCatCPUTrees}{200}
\newcommand{\ComparisonCatRerunTrial}{19}
\newcommand{\ComparisonCatSourceTrial}{1}
\newcommand{\ComparisonConfigurations}{20}
\newcommand{\ComparisonLightGBMTestRawAP}{0.328}
\newcommand{\ComparisonSeed}{7}
\newcommand{\ComparisonTestPositives}{61,119}
\newcommand{\ComparisonXGBoostCalibAP}{0.310}
\newcommand{\ComparisonXGBoostGain}{-0.0057}
\newcommand{\ComparisonXGBoostGainHigh}{0.0002}
\newcommand{\ComparisonXGBoostGainLow}{-0.0100}
\newcommand{\ComparisonXGBoostSelectAP}{0.345}
\newcommand{\ComparisonXGBoostTestCalibratedAP}{0.319}
\newcommand{\ComparisonXGBoostTestRawAP}{0.324}
\newcommand{\ConfidencePercent}{95}
\newcommand{\CoordinateSnap}{1e-9}
\newcommand{\DggydfBurnedArea}{12,658,157}
\newcommand{\DggydfFireYear}{2024}
\newcommand{\DggydfPlanYear}{2026}
\newcommand{\DryThreshold}{1}
\newcommand{\ECEBins}{10}
\newcommand{\EarlyStopRounds}{30}
\newcommand{\EligibleBlocks}{124}
\newcommand{\EligibleCells}{37,146}
\newcommand{\ExampleDetections}{131,310}
\newcommand{\ExampleEnd}{2024-09-09}
\newcommand{\ExampleIssue}{2024-09-02}
\newcommand{\ExamplePositiveCells}{3,352}
\newcommand{\ExcludedArea}{10780.55}
\newcommand{\ExclusionCalibrationEligible}{1,180,240}
\newcommand{\ExclusionCalibrationGeography}{359,080}
\newcommand{\ExclusionCalibrationHistory}{206,542}
\newcommand{\ExclusionCalibrationMissing}{148,584}
\newcommand{\ExclusionCalibrationPurged}{111,438}
\newcommand{\ExclusionCalibrationScanned}{2,005,884}
\newcommand{\ExclusionGeographicEligible}{809,840}
\newcommand{\ExclusionGeographicGeography}{4,189,852}
\newcommand{\ExclusionGeographicHistory}{275,040}
\newcommand{\ExclusionGeographicMissing}{557,190}
\newcommand{\ExclusionGeographicPurged}{111,438}
\newcommand{\ExclusionGeographicScanned}{5,943,360}
\newcommand{\ExclusionSelectionEligible}{1,357,276}
\newcommand{\ExclusionSelectionGeography}{382,000}
\newcommand{\ExclusionSelectionHistory}{118,024}
\newcommand{\ExclusionSelectionMissing}{37,146}
\newcommand{\ExclusionSelectionPurged}{111,438}
\newcommand{\ExclusionSelectionScanned}{2,005,884}
\newcommand{\ExclusionTemporalEligible}{3,127,636}
\newcommand{\ExclusionTemporalGeography}{1,084,880}
\newcommand{\ExclusionTemporalHistory}{1,062,216}
\newcommand{\ExclusionTemporalMissing}{557,190}
\newcommand{\ExclusionTemporalPurged}{111,438}
\newcommand{\ExclusionTemporalScanned}{5,943,360}
\newcommand{\ExclusionTrainEligible}{11,979,436}
\newcommand{\ExclusionTrainGeography}{3,162,960}
\newcommand{\ExclusionTrainHistory}{236,048}
\newcommand{\ExclusionTrainMissing}{74,292}
\newcommand{\ExclusionTrainPurged}{111,438}
\newcommand{\ExclusionTrainScanned}{15,564,174}
\newcommand{\ExtensionCompositeDays}{16}
\newcommand{\ExtensionCurrentAge}{64}
\newcommand{\ExtensionDiagnosticBHigh}{0.0111}
\newcommand{\ExtensionDiagnosticBLow}{0.0006}
\newcommand{\ExtensionDiagnosticBRows}{580,563}
\newcommand{\ExtensionDiagnosticEHigh}{0.0131}
\newcommand{\ExtensionDiagnosticELow}{-0.0000}
\newcommand{\ExtensionDiagnosticERows}{576,222}
\newcommand{\ExtensionDiagnosticFHigh}{0.0103}
\newcommand{\ExtensionDiagnosticFLow}{-0.0074}
\newcommand{\ExtensionDiagnosticFRows}{684,528}
\newcommand{\ExtensionDiagnosticGHigh}{0.0106}
\newcommand{\ExtensionDiagnosticGLow}{-0.0027}
\newcommand{\ExtensionDiagnosticGRows}{607,980}
\newcommand{\ExtensionDiagnosticHHigh}{0.0115}
\newcommand{\ExtensionDiagnosticHLow}{0.0030}
\newcommand{\ExtensionDiagnosticHRows}{1,172,000}
\newcommand{\ExtensionDiagnosticIHigh}{0.0195}
\newcommand{\ExtensionDiagnosticILow}{-0.0011}
\newcommand{\ExtensionDiagnosticIRows}{679,634}
\newcommand{\ExtensionDiagnosticJHigh}{0.0205}
\newcommand{\ExtensionDiagnosticJLow}{0.0010}
\newcommand{\ExtensionDiagnosticJRows}{603,525}
\newcommand{\ExtensionHFive}{H5}
\newcommand{\ExtensionHFour}{H4}
\newcommand{\ExtensionHThree}{H3}
\newcommand{\ExtensionLag}{16}
\newcommand{\ExtensionLandAge}{2}
\newcommand{\ExtensionLandCoverage}{0.95}
\newcommand{\ExtensionLandProduct}{MapBiomas Bolivia Collection 4}
\newcommand{\ExtensionLongLag}{32}
\newcommand{\ExtensionProduct}{MOD13A2.061}
\newcommand{\ExtensionReferenceAge}{128}
\newcommand{\ExtensionReferenceYears}{4}
\newcommand{\ExtensionRunBLag}{32}
\newcommand{\ExtensionRunELag}{32}
\newcommand{\ExtensionRunFLag}{16}
\newcommand{\ExtensionRunGLag}{32}
\newcommand{\ExtensionRunILag}{16}
\newcommand{\ExtensionRunJLag}{32}
\newcommand{\ExtensionTestAHigh}{0.0128}
\newcommand{\ExtensionTestALow}{-0.0033}
\newcommand{\ExtensionTestARows}{1,754,909}
\newcommand{\ExtensionTestBHigh}{0.0040}
\newcommand{\ExtensionTestBLow}{0.0003}
\newcommand{\ExtensionTestBRows}{3,105,800}
\newcommand{\ExtensionTestCHigh}{0.0164}
\newcommand{\ExtensionTestCLow}{0.0020}
\newcommand{\ExtensionTestCRows}{1,742,193}
\newcommand{\ExtensionValidArea}{0.5}
\newcommand{\ExtensionValidationADecision}{Inconclusive}
\newcommand{\ExtensionValidationAHigh}{0.0174}
\newcommand{\ExtensionValidationALow}{-0.0083}
\newcommand{\ExtensionValidationARows}{645,548}
\newcommand{\ExtensionValidationBDecision}{Inconclusive}
\newcommand{\ExtensionValidationBHigh}{0.0039}
\newcommand{\ExtensionValidationBLow}{-0.0046}
\newcommand{\ExtensionValidationBRows}{1,172,000}
\newcommand{\ExtensionValidationCDecision}{Inconclusive}
\newcommand{\ExtensionValidationCHigh}{0.0117}
\newcommand{\ExtensionValidationCLow}{-0.0008}
\newcommand{\ExtensionValidationCRows}{640,862}
\newcommand{\FanBurnedArea}{5,358,492}
\newcommand{\FanReportDate}{25 September 2019}
\newcommand{\FirstHypothesis}{1}
\newcommand{\FirstIssue}{2013-01-07}
\newcommand{\GAPGain}{0.2032}
\newcommand{\GAPGainHigh}{0.2325}
\newcommand{\GAPGainLow}{0.1454}
\newcommand{\GBlocks}{24}
\newcommand{\GBoostAP}{0.384}
\newcommand{\GBoostAPHigh}{0.468}
\newcommand{\GBoostAPLow}{0.250}
\newcommand{\GBoostAUC}{0.944}
\newcommand{\GBoostBrier}{0.011}
\newcommand{\GBoostECE}{0.00092}
\newcommand{\GBoostLogLoss}{0.043}
\newcommand{\GBoostRecallLarge}{0.673}
\newcommand{\GBoostRecallSmall}{0.469}
\newcommand{\GBoostRows}{809,840}
\newcommand{\GClimatologyAP}{0.071}
\newcommand{\GClimatologyAPHigh}{0.135}
\newcommand{\GClimatologyAPLow}{0.031}
\newcommand{\GClimatologyAUC}{0.850}
\newcommand{\GClimatologyBrier}{0.013}
\newcommand{\GClimatologyECE}{0.0023}
\newcommand{\GClimatologyLogLoss}{0.061}
\newcommand{\GClimatologyRecallLarge}{0.122}
\newcommand{\GClimatologyRecallSmall}{0.104}
\newcommand{\GLogisticAP}{0.351}
\newcommand{\GLogisticAPHigh}{0.436}
\newcommand{\GLogisticAPLow}{0.218}
\newcommand{\GLogisticAUC}{0.940}
\newcommand{\GLogisticBrier}{0.011}
\newcommand{\GLogisticECE}{0.0028}
\newcommand{\GLogisticLogLoss}{0.046}
\newcommand{\GLogisticRecallLarge}{0.651}
\newcommand{\GLogisticRecallSmall}{0.453}
\newcommand{\GPersistenceAP}{0.181}
\newcommand{\GPersistenceAPHigh}{0.243}
\newcommand{\GPersistenceAPLow}{0.102}
\newcommand{\GPersistenceAUC}{0.672}
\newcommand{\GPersistenceBrier}{0.012}
\newcommand{\GPersistenceECE}{0.0096}
\newcommand{\GPersistenceLogLoss}{0.062}
\newcommand{\GPersistenceRecallLarge}{0.361}
\newcommand{\GPersistenceRecallSmall}{0.300}
\newcommand{\GeoBoostCalibrationAP}{0.310}
\newcommand{\GeoBoostCalibrationBrier}{0.021}
\newcommand{\GeoBoostCalibrationECE}{0.000070}
\newcommand{\GeoBoostSelectionAP}{0.326}
\newcommand{\GeoClimatologyCalibrationAP}{0.076}
\newcommand{\GeoClimatologyCalibrationBrier}{0.025}
\newcommand{\GeoClimatologyCalibrationECE}{0.0024}
\newcommand{\GeoClimatologySelectionAP}{0.083}
\newcommand{\GeoLogisticCalibrationAP}{0.270}
\newcommand{\GeoLogisticCalibrationBrier}{0.022}
\newcommand{\GeoLogisticCalibrationECE}{0.00011}
\newcommand{\GeoLogisticSelectionAP}{0.293}
\newcommand{\GeoPersistenceCalibrationAP}{0.150}
\newcommand{\GeoPersistenceCalibrationBrier}{0.023}
\newcommand{\GeoPersistenceCalibrationECE}{0.014}
\newcommand{\GeoPersistenceSelectionAP}{0.148}
\newcommand{\GridDegrees}{0.05}
\newcommand{\HOneHigh}{0.2179}
\newcommand{\HOneLow}{0.0998}
\newcommand{\HTwoHigh}{0.0019}
\newcommand{\HTwoLow}{-0.0041}
\newcommand{\HeldOutArea}{219395.59}
\newcommand{\HeldOutBlocks}{24}
\newcommand{\HeldOutCells}{7,640}
\newcommand{\HistogramBins}{63}
\newcommand{\HistoryDays}{28}
\newcommand{\HoldoutPercent}{20}
\newcommand{\HoldoutSeed}{20261031}
\newcommand{\HorizonDays}{7}
\newcommand{\IssueWeeks}{678}
\newcommand{\LandPilotCells}{1,000}
\newcommand{\LandPilotCoverageError}{0.0002452}
\newcommand{\LandPilotFractionError}{0.0006534}
\newcommand{\LandPilotTolerance}{0.01}
\newcommand{\LargeBudget}{10}
\newcommand{\LastIssue}{2025-12-29}
\newcommand{\LastWeekIndex}{51}
\newcommand{\LogClip}{1e-15}
\newcommand{\LongHistoryDays}{365}
\newcommand{\LowerClip}{1e-6}
\newcommand{\ModelThreads}{4}
\newcommand{\NationalArea}{1068783.84}
\newcommand{\NegativePercent}{10}
\newcommand{\One}{1}
\newcommand{\OpeningCount}{9}
\newcommand{\OutageHours}{6}
\newcommand{\ParameterABaseline}{climatology}
\newcommand{\ParameterAFeatureCount}{21}
\newcommand{\ParameterAModel}{lightgbm\_08}
\newcommand{\ParameterBBaseline}{persistence}
\newcommand{\ParameterBFeatureCount}{20}
\newcommand{\ParameterBModel}{lightgbm\_04}
\newcommand{\ParameterCBaseline}{climatology}
\newcommand{\ParameterCFeatureCount}{26}
\newcommand{\ParameterCModel}{lightgbm\_16}
\newcommand{\ParameterDBaseline}{persistence}
\newcommand{\ParameterDFeatureCount}{25}
\newcommand{\ParameterDModel}{lightgbm\_04}
\newcommand{\ProtocolVersion}{1.1}
\newcommand{\PurgedHeldPositivePercent}{0.62}
\newcommand{\PurgedLabelPercent}{100.00}
\newcommand{\PurgedPositivePercent}{0.64}
\newcommand{\PurgedRows}{111,438}
\newcommand{\RainLongDays}{90}
\newcommand{\RainMiddleDays}{30}
\newcommand{\RainShortDays}{7}
\newcommand{\RainfallAUCLower}{-0.0012685}
\newcommand{\RainfallAUCUpper}{-0.0003378}
\newcommand{\RainfallBrierLower}{-0.0000463}
\newcommand{\RainfallBrierUpper}{-0.0000055}
\newcommand{\RainfallECELower}{-0.0010025}
\newcommand{\RainfallECEUpper}{-0.0000100}
\newcommand{\SantaCruzBoostAP}{0.381}
\newcommand{\SantaCruzBoostBrier}{0.018}
\newcommand{\SantaCruzBoostECE}{0.0038}
\newcommand{\SantaCruzBoostRecallLarge}{0.573}
\newcommand{\SantaCruzBoostRecallSmall}{0.388}
\newcommand{\SantaCruzClimatologyAP}{0.161}
\newcommand{\SantaCruzClimatologyBrier}{0.022}
\newcommand{\SantaCruzClimatologyECE}{0.012}
\newcommand{\SantaCruzClimatologyRecallLarge}{0.353}
\newcommand{\SantaCruzClimatologyRecallSmall}{0.216}
\newcommand{\SantaCruzClimatologyRows}{1,095,934}
\newcommand{\SantaCruzLogisticAP}{0.333}
\newcommand{\SantaCruzLogisticBrier}{0.019}
\newcommand{\SantaCruzLogisticECE}{0.0045}
\newcommand{\SantaCruzLogisticRecallLarge}{0.558}
\newcommand{\SantaCruzLogisticRecallSmall}{0.373}
\newcommand{\SantaCruzPersistenceAP}{0.202}
\newcommand{\SantaCruzPersistenceBrier}{0.020}
\newcommand{\SantaCruzPersistenceECE}{0.013}
\newcommand{\SantaCruzPersistenceRecallLarge}{0.367}
\newcommand{\SantaCruzPersistenceRecallSmall}{0.295}
\newcommand{\SearchCount}{20}
\newcommand{\SearchFractions}{0.7, 0.85, 1}
\newcommand{\SearchLeaves}{7, 15, 31, 63}
\newcommand{\SearchMinimumLeaves}{50, 200, 1,000}
\newcommand{\SearchPenalties}{0, 1, 10}
\newcommand{\SearchRates}{0.03, 0.06, 0.1}
\newcommand{\SearchSeed}{7}
\newcommand{\SecondHypothesis}{2}
\newcommand{\SelectHeldPositivePercent}{1.63}
\newcommand{\SelectLabelPercent}{98.04}
\newcommand{\SelectPositivePercent}{2.38}
\newcommand{\SelectRows}{1,894,446}
\newcommand{\SelectYear}{2021}
\newcommand{\SelectedDetections}{3,614,079}
\newcommand{\SensitivityAGAPHigh}{0.473}
\newcommand{\SensitivityAGAPLow}{0.251}
\newcommand{\SensitivityAGGainHigh}{0.2340}
\newcommand{\SensitivityAGGainLow}{0.1427}
\newcommand{\SensitivityAGGainPoint}{0.2036}
\newcommand{\SensitivityAGPoint}{0.385}
\newcommand{\SensitivityATAPHigh}{0.391}
\newcommand{\SensitivityATAPLow}{0.229}
\newcommand{\SensitivityATGainHigh}{0.2225}
\newcommand{\SensitivityATGainLow}{0.1059}
\newcommand{\SensitivityATGainPoint}{0.1618}
\newcommand{\SensitivityATPoint}{0.332}
\newcommand{\SensitivityBGAPHigh}{0.472}
\newcommand{\SensitivityBGAPLow}{0.252}
\newcommand{\SensitivityBGGainHigh}{0.2354}
\newcommand{\SensitivityBGGainLow}{0.1488}
\newcommand{\SensitivityBGGainPoint}{0.2066}
\newcommand{\SensitivityBGPoint}{0.387}
\newcommand{\SensitivityBTAPHigh}{0.375}
\newcommand{\SensitivityBTAPLow}{0.215}
\newcommand{\SensitivityBTGainHigh}{0.2206}
\newcommand{\SensitivityBTGainLow}{0.0908}
\newcommand{\SensitivityBTGainPoint}{0.1493}
\newcommand{\SensitivityBTPoint}{0.316}
\newcommand{\SensitivityCGAPHigh}{0.476}
\newcommand{\SensitivityCGAPLow}{0.252}
\newcommand{\SensitivityCGGainHigh}{0.2398}
\newcommand{\SensitivityCGGainLow}{0.1492}
\newcommand{\SensitivityCGGainPoint}{0.2083}
\newcommand{\SensitivityCGPoint}{0.389}
\newcommand{\SensitivityCTAPHigh}{0.386}
\newcommand{\SensitivityCTAPLow}{0.228}
\newcommand{\SensitivityCTGainHigh}{0.2208}
\newcommand{\SensitivityCTGainLow}{0.1057}
\newcommand{\SensitivityCTGainPoint}{0.1603}
\newcommand{\SensitivityCTPoint}{0.327}
\newcommand{\SensitivityDGAPHigh}{0.438}
\newcommand{\SensitivityDGAPLow}{0.214}
\newcommand{\SensitivityDGGainHigh}{0.2047}
\newcommand{\SensitivityDGGainLow}{0.1071}
\newcommand{\SensitivityDGGainPoint}{0.1673}
\newcommand{\SensitivityDGPoint}{0.348}
\newcommand{\SensitivityDTAPHigh}{0.364}
\newcommand{\SensitivityDTAPLow}{0.216}
\newcommand{\SensitivityDTGainHigh}{0.1984}
\newcommand{\SensitivityDTGainLow}{0.0952}
\newcommand{\SensitivityDTGainPoint}{0.1434}
\newcommand{\SensitivityDTPoint}{0.310}
\newcommand{\SensitivityReportCount}{5}
\newcommand{\SmallBudget}{5}
\newcommand{\SmoothingWeeks}{2}
\newcommand{\StrictBoostCalibrationAP}{0.316}
\newcommand{\StrictBoostCalibrationBrier}{0.021}
\newcommand{\StrictBoostCalibrationECE}{0.000071}
\newcommand{\StrictBoostSelectionAP}{0.347}
\newcommand{\StrictClimatologyCalibrationAP}{0.181}
\newcommand{\StrictClimatologyCalibrationBrier}{0.023}
\newcommand{\StrictClimatologyCalibrationECE}{0.0033}
\newcommand{\StrictClimatologySelectionAP}{0.219}
\newcommand{\StrictLogisticCalibrationAP}{0.283}
\newcommand{\StrictLogisticCalibrationBrier}{0.022}
\newcommand{\StrictLogisticCalibrationECE}{0.00011}
\newcommand{\StrictLogisticSelectionAP}{0.326}
\newcommand{\StrictPersistenceCalibrationAP}{0.150}
\newcommand{\StrictPersistenceCalibrationBrier}{0.023}
\newcommand{\StrictPersistenceCalibrationECE}{0.014}
\newcommand{\StrictPersistenceSelectionAP}{0.148}
\newcommand{\SupplementCNNBatchWeeks}{4}
\newcommand{\SupplementCNNBest}{12}
\newcommand{\SupplementCNNDepth}{2}
\newcommand{\SupplementCNNDropout}{0}
\newcommand{\SupplementCNNEpochs}{12}
\newcommand{\SupplementCNNLearningRate}{0.003}
\newcommand{\SupplementCNNPatience}{3}
\newcommand{\SupplementCNNTrial}{3}
\newcommand{\SupplementCNNWeightDecay}{0.0001}
\newcommand{\SupplementCNNWidth}{24}
\newcommand{\SupplementCatBaggingTemperature}{0}
\newcommand{\SupplementCatBest}{200}
\newcommand{\SupplementCatBootstrapType}{Bayesian}
\newcommand{\SupplementCatBorderCount}{63}
\newcommand{\SupplementCatCheckpoint}{10}
\newcommand{\SupplementCatDepth}{6}
\newcommand{\SupplementCatIterations}{400}
\newcommand{\SupplementCatLLeafReg}{10}
\newcommand{\SupplementCatLearningRate}{0.06}
\newcommand{\SupplementCatLossFunction}{Logloss}
\newcommand{\SupplementCatRandomSeed}{7}
\newcommand{\SupplementCatRandomStrength}{1}
\newcommand{\SupplementCatTaskType}{GPU}
\newcommand{\SupplementCatThreadCount}{4}
\newcommand{\SupplementCatTrial}{19}
\newcommand{\SupplementCodeCommit}{81cfdf88b134d8d9a416abd904402020d2d50535}
\newcommand{\SupplementDaysFortnight}{14}
\newcommand{\SupplementDaysMonth}{28}
\newcommand{\SupplementDaysQuarter}{91}
\newcommand{\SupplementDaysShort}{7}
\newcommand{\SupplementDaysYear}{365}
\newcommand{\SupplementFeatureAName}{count\_last\_7d}
\newcommand{\SupplementFeatureBName}{count\_last\_14d}
\newcommand{\SupplementFeatureCName}{count\_last\_28d}
\newcommand{\SupplementFeatureDName}{count\_last\_91d}
\newcommand{\SupplementFeatureEName}{count\_last\_365d}
\newcommand{\SupplementFeatureFName}{any\_last\_7d}
\newcommand{\SupplementFeatureGName}{days\_since\_last}
\newcommand{\SupplementFeatureHName}{no\_prior\_detection}
\newcommand{\SupplementFeatureIName}{frp\_sum\_28d}
\newcommand{\SupplementFeatureJName}{same\_week\_last\_year}
\newcommand{\SupplementFeatureKName}{neighborhood\_3x3\_last\_7d}
\newcommand{\SupplementFeatureLName}{neighborhood\_7x7\_last\_7d}
\newcommand{\SupplementFeatureMName}{neighborhood\_3x3\_last\_28d}
\newcommand{\SupplementFeatureNName}{neighborhood\_7x7\_last\_28d}
\newcommand{\SupplementFeatureOName}{national\_last\_7d}
\newcommand{\SupplementFeaturePName}{national\_last\_28d}
\newcommand{\SupplementFeatureQName}{week\_of\_year\_sin}
\newcommand{\SupplementFeatureRName}{week\_of\_year\_cos}
\newcommand{\SupplementFeatureSName}{train\_climatology\_cell}
\newcommand{\SupplementFeatureTName}{train\_climatology\_dept}
\newcommand{\SupplementFeatureUName}{window\_obs\_complete\_365d}
\newcommand{\SupplementGeographicBestIteration}{36}
\newcommand{\SupplementGeographicFeatureFraction}{0.7}
\newcommand{\SupplementGeographicLambdaL}{10}
\newcommand{\SupplementGeographicLearningRate}{0.03}
\newcommand{\SupplementGeographicMinDataInLeaf}{200}
\newcommand{\SupplementGeographicNumLeaves}{63}
\newcommand{\SupplementGradientClip}{5}
\newcommand{\SupplementLogisticFraction}{0.1}
\newcommand{\SupplementLogisticIterations}{1,000}
\newcommand{\SupplementLogisticSampledRows}{1,427,749}
\newcommand{\SupplementLogisticSolver}{lbfgs}
\newcommand{\SupplementLogisticWeight}{10}
\newcommand{\SupplementMaskMetres}{375}
\newcommand{\SupplementNeighborhoodLarge}{7}
\newcommand{\SupplementNeighborhoodSmall}{3}
\newcommand{\SupplementPersistenceHigh}{5}
\newcommand{\SupplementPersistenceUpper}{4}
\newcommand{\SupplementPointKernel}{1}
\newcommand{\SupplementPrevalence}{1.95}
\newcommand{\SupplementPreviousEnd}{357}
\newcommand{\SupplementPreviousStart}{364}
\newcommand{\SupplementRainfallRows}{3,127,636}
\newcommand{\SupplementRecencyCap}{730}
\newcommand{\SupplementRetainedWeeks}{106}
\newcommand{\SupplementSearchCatboostBaggingTemperature}{0, 0.5, 1}
\newcommand{\SupplementSearchCatboostDepth}{4, 5, 6, 7}
\newcommand{\SupplementSearchCatboostLLeafReg}{1, 10, 50}
\newcommand{\SupplementSearchCatboostLearningRate}{0.03, 0.06, 0.1}
\newcommand{\SupplementSearchCatboostRandomStrength}{0, 1}
\newcommand{\SupplementSearchSpatialcnnDepth}{2, 3}
\newcommand{\SupplementSearchSpatialcnnDropout}{0, 0.1}
\newcommand{\SupplementSearchSpatialcnnLearningRate}{0.001, 0.003}
\newcommand{\SupplementSearchSpatialcnnWeightDecay}{0.0001, 0.001}
\newcommand{\SupplementSearchSpatialcnnWidth}{8, 16, 24}
\newcommand{\SupplementSearchXgboostColsampleBytree}{0.7, 0.85, 1}
\newcommand{\SupplementSearchXgboostEta}{0.03, 0.06, 0.1}
\newcommand{\SupplementSearchXgboostLambda}{1, 10, 100}
\newcommand{\SupplementSearchXgboostMaxDepth}{3, 4, 5, 6}
\newcommand{\SupplementSearchXgboostMinChildWeight}{10, 50, 200}
\newcommand{\SupplementSklearnVersion}{1.9.1}
\newcommand{\SupplementSpatialKernel}{3}
\newcommand{\SupplementTemporalBestIteration}{78}
\newcommand{\SupplementTemporalFeatureFraction}{0.7}
\newcommand{\SupplementTemporalLambdaL}{10}
\newcommand{\SupplementTemporalLearningRate}{0.03}
\newcommand{\SupplementTemporalMinDataInLeaf}{1,000}
\newcommand{\SupplementTemporalNumLeaves}{15}
\newcommand{\SupplementTestWeeks}{157}
\newcommand{\SupplementXGBBest}{65}
\newcommand{\SupplementXGBColsampleBytree}{0.85}
\newcommand{\SupplementXGBDevice}{cuda}
\newcommand{\SupplementXGBEta}{0.03}
\newcommand{\SupplementXGBLambda}{1}
\newcommand{\SupplementXGBMaxBin}{63}
\newcommand{\SupplementXGBMaxDepth}{5}
\newcommand{\SupplementXGBMinChildWeight}{10}
\newcommand{\SupplementXGBNthread}{4}
\newcommand{\SupplementXGBObjective}{binary:logistic}
\newcommand{\SupplementXGBSeed}{7}
\newcommand{\SupplementXGBSubsample}{1}
\newcommand{\SupplementXGBTreeMethod}{hist}
\newcommand{\SupplementXGBTrial}{9}
\newcommand{\SupplementYearAWeeks}{42}
\newcommand{\SupplementYearBWeeks}{23}
\newcommand{\SupplementYearCWeeks}{41}
\newcommand{\TAPGain}{0.1567}
\newcommand{\TBlocks}{28}
\newcommand{\TBoostAP}{0.323}
\newcommand{\TBoostAPHigh}{0.383}
\newcommand{\TBoostAPLow}{0.222}
\newcommand{\TBoostAUC}{0.928}
\newcommand{\TBoostBrier}{0.016}
\newcommand{\TBoostECE}{0.0031}
\newcommand{\TBoostLogLoss}{0.062}
\newcommand{\TBoostRecallLarge}{0.558}
\newcommand{\TBoostRecallLargePercent}{55.8}
\newcommand{\TBoostRecallSmall}{0.384}
\newcommand{\TBoostRecallSmallPercent}{38.4}
\newcommand{\TBoostRows}{3,127,636}
\newcommand{\TClimatologyAP}{0.167}
\newcommand{\TClimatologyAPHigh}{0.214}
\newcommand{\TClimatologyAPLow}{0.105}
\newcommand{\TClimatologyAUC}{0.830}
\newcommand{\TClimatologyBrier}{0.018}
\newcommand{\TClimatologyECE}{0.0070}
\newcommand{\TClimatologyLogLoss}{0.108}
\newcommand{\TClimatologyRecallLarge}{0.397}
\newcommand{\TClimatologyRecallLargePercent}{39.7}
\newcommand{\TClimatologyRecallSmall}{0.251}
\newcommand{\TLogisticAP}{0.293}
\newcommand{\TLogisticAPHigh}{0.353}
\newcommand{\TLogisticAPLow}{0.195}
\newcommand{\TLogisticAUC}{0.923}
\newcommand{\TLogisticBrier}{0.016}
\newcommand{\TLogisticECE}{0.0039}
\newcommand{\TLogisticLogLoss}{0.064}
\newcommand{\TLogisticRecallLarge}{0.545}
\newcommand{\TLogisticRecallSmall}{0.370}
\newcommand{\TPersistenceAP}{0.143}
\newcommand{\TPersistenceAPHigh}{0.187}
\newcommand{\TPersistenceAPLow}{0.085}
\newcommand{\TPersistenceAUC}{0.648}
\newcommand{\TPersistenceBrier}{0.017}
\newcommand{\TPersistenceECE}{0.011}
\newcommand{\TPersistenceLogLoss}{0.083}
\newcommand{\TPersistenceRecallLarge}{0.323}
\newcommand{\TPersistenceRecallSmall}{0.265}
\newcommand{\TestHeldPositivePercent}{1.71}
\newcommand{\TestLabelPercent}{90.45}
\newcommand{\TestPositivePercent}{2.46}
\newcommand{\TestRows}{5,831,922}
\newcommand{\TestYearRange}{2023--2025}
\newcommand{\TotalRows}{25,184,988}
\newcommand{\TrainHeldPositivePercent}{1.53}
\newcommand{\TrainLabelPercent}{99.52}
\newcommand{\TrainPositivePercent}{2.18}
\newcommand{\TrainRows}{15,452,736}
\newcommand{\Two}{2}
\newcommand{\UnobservedDays}{64}
\newcommand{\YearA}{2023}
\newcommand{\YearABoostAP}{0.347}
\newcommand{\YearABoostBrier}{0.019}
\newcommand{\YearABoostECE}{0.0013}
\newcommand{\YearABoostRecallLarge}{0.543}
\newcommand{\YearABoostRecallSmall}{0.367}
\newcommand{\YearAClimatologyAP}{0.209}
\newcommand{\YearAClimatologyBrier}{0.021}
\newcommand{\YearAClimatologyECE}{0.0054}
\newcommand{\YearAClimatologyRecallLarge}{0.402}
\newcommand{\YearAClimatologyRecallSmall}{0.254}
\newcommand{\YearAClimatologyRows}{1,239,252}
\newcommand{\YearALogisticAP}{0.332}
\newcommand{\YearALogisticBrier}{0.019}
\newcommand{\YearALogisticECE}{0.0023}
\newcommand{\YearALogisticRecallLarge}{0.533}
\newcommand{\YearALogisticRecallSmall}{0.357}
\newcommand{\YearAPersistenceAP}{0.156}
\newcommand{\YearAPersistenceBrier}{0.021}
\newcommand{\YearAPersistenceECE}{0.015}
\newcommand{\YearAPersistenceRecallLarge}{0.331}
\newcommand{\YearAPersistenceRecallSmall}{0.267}
\newcommand{\YearB}{2024}
\newcommand{\YearBBoostAP}{0.374}
\newcommand{\YearBBoostBrier}{0.014}
\newcommand{\YearBBoostECE}{0.0018}
\newcommand{\YearBBoostRecallLarge}{0.588}
\newcommand{\YearBBoostRecallSmall}{0.433}
\newcommand{\YearBClimatologyAP}{0.100}
\newcommand{\YearBClimatologyBrier}{0.017}
\newcommand{\YearBClimatologyECE}{0.0088}
\newcommand{\YearBClimatologyRecallLarge}{0.309}
\newcommand{\YearBClimatologyRecallSmall}{0.189}
\newcommand{\YearBClimatologyRows}{678,638}
\newcommand{\YearBLogisticAP}{0.319}
\newcommand{\YearBLogisticBrier}{0.015}
\newcommand{\YearBLogisticECE}{0.0021}
\newcommand{\YearBLogisticRecallLarge}{0.568}
\newcommand{\YearBLogisticRecallSmall}{0.407}
\newcommand{\YearBPersistenceAP}{0.208}
\newcommand{\YearBPersistenceBrier}{0.015}
\newcommand{\YearBPersistenceECE}{0.011}
\newcommand{\YearBPersistenceRecallLarge}{0.400}
\newcommand{\YearBPersistenceRecallSmall}{0.302}
\newcommand{\YearC}{2025}
\newcommand{\YearCBoostAP}{0.246}
\newcommand{\YearCBoostBrier}{0.013}
\newcommand{\YearCBoostECE}{0.0059}
\newcommand{\YearCBoostRecallLarge}{0.562}
\newcommand{\YearCBoostRecallSmall}{0.380}
\newcommand{\YearCClimatologyAP}{0.166}
\newcommand{\YearCClimatologyBrier}{0.014}
\newcommand{\YearCClimatologyECE}{0.010}
\newcommand{\YearCClimatologyRecallLarge}{0.448}
\newcommand{\YearCClimatologyRecallSmall}{0.289}
\newcommand{\YearCClimatologyRows}{1,209,746}
\newcommand{\YearCLogisticAP}{0.229}
\newcommand{\YearCLogisticBrier}{0.013}
\newcommand{\YearCLogisticECE}{0.0075}
\newcommand{\YearCLogisticRecallLarge}{0.549}
\newcommand{\YearCLogisticRecallSmall}{0.366}
\newcommand{\YearCPersistenceAP}{0.084}
\newcommand{\YearCPersistenceBrier}{0.014}
\newcommand{\YearCPersistenceECE}{0.013}
\newcommand{\YearCPersistenceRecallLarge}{0.257}
\newcommand{\YearCPersistenceRecallSmall}{0.238}
\newcommand{\Zero}{0}

\title{Predicting next-week VIIRS fire detections in Bolivia: a reproducible benchmark with temporal and geographic evaluation}
\author{Victor Hugo Cuellar Flores\\[3pt]\small\textit{Santa Cruz de la Sierra, Bolivia}\\[3pt]\small\href{mailto:cvictorhugo39@hotmail.com}{cvictorhugo39@hotmail.com}}
\hypersetup{pdfauthor={Victor Hugo Cuellar Flores},pdftitle={Predicting next-week VIIRS fire detections in Bolivia: a reproducible benchmark with temporal and geographic evaluation}}
\date{}
\begin{document}
\maketitle
\vspace{-\baselineskip}
\begin{abstract}
Weekly detection forecasts could help prioritize fire monitoring across Bolivia. We estimate the probability of at least one fire detection by the Visible Infrared Imaging Radiometer Suite (VIIRS) on the Suomi National Polar-orbiting Partnership satellite in each \num{\GridDegrees}-degree cell during the following \HorizonDays{} days. Primary models were fixed before test scoring. Calibrated LightGBM, a gradient-boosted decision-tree model, reaches \TestYearRange{} temporal-test average precision (AP) of \TBoostAP{} (\ConfidencePercent{}\% block-bootstrap interval \TBoostAPLow{}--\TBoostAPHigh{}), compared with \TClimatologyAP{} for climatology and \TPersistenceAP{} for persistence. Its paired AP gain over climatology is \TAPGain{} (\HOneLow{}--\HOneHigh{}). A separately selected geographic model reaches AP \GBoostAP{} versus \GPersistenceAP{} for persistence in regions excluded from fitting. In the temporal-test population, reviewing the highest-ranked \SmallBudget{}\% and \LargeBudget{}\% of eligible area would have covered \TBoostRecallSmallPercent{}\% and \TBoostRecallLargePercent{}\% of cell-weeks with detections. Probabilities still overestimate detection frequency in the highest-probability group. Retrospective rainfall provides no clear AP improvement; the primary validation comparisons for vegetation, land cover and their combination are inconclusive. Positive secondary or exploratory intervals do not change those decisions. Evaluation requires complete target and recent-history observations and measures satellite detections, not individual fires, causes or burned area. Prospective evaluation should archive inputs at each weekly issue and measure ranking, calibration and the usefulness of selected cells for monitoring decisions.
\end{abstract}
\noindent\textbf{Keywords:} VIIRS active fire; fire detection prediction; Bolivia; benchmark; probability calibration; temporal and spatial validation.

\section{Introduction}
In the Chiquitano dry forest, chaqueo uses fire to prepare land for agriculture and pasture \citep{maillard2024}, within a broader lowland land-conversion context involving mechanized agriculture, cattle ranching and small-scale farming \citep{muller2012}. \citet{fan20190925} reported a preliminary national burned area of \num{\FanBurnedArea}\,ha as of \FanReportDate{}. Separately, \citet{dggydf2026} lists a national burned area of \num{\DggydfBurnedArea}\,ha for \DggydfFireYear{} in its \DggydfPlanYear{} prevention plan. August and September were the peak fire months in the Chiquitania hotspot record \citep{devisscher2016}. This seasonal concentration motivates short-horizon forecasts to prioritize monitoring, while satellite detections alone reveal neither a fire's cause nor its legality.

We test whether earlier satellite detections improve next-week rankings over training-only climatology and persistence in future years and held-out regions. Protocol \ProtocolVersion{} freezes the following hypotheses.

\noindent\textbf{H\FirstHypothesis{}.} A model using only information available at issue time (strict variant) improves average precision over the better of climatology and persistence on the \TestYearRange{} test years.

\noindent\textbf{H\SecondHypothesis{}.} Adding retrospective covariates (CHIRPS v\ChirpsMajor{} rainfall, MapBiomas land cover) changes test average precision by less than the block-bootstrap uncertainty of the strict model.

The registered operational rule supports H\SecondHypothesis{} when the paired \ConfidencePercent{}\% calendar-month-block bootstrap interval for temporal-test AP(retrospective) minus AP(strict) contains zero; this is not an equivalence test, and land cover was unavailable to the primary comparison and is evaluated separately as \ExtensionHFour{}.

\section{Related work}
Bolivian studies address spatial wildfire susceptibility in Santa Cruz \citep{bustillo2021}, climate and development interactions in the Chiquitania \citep{devisscher2016}, regional fire impacts \citep{romero2019}, and post-fire phenology and regeneration \citep{maillard2022}.

South American fire-season forecasts use sea-surface temperature anomalies \citep{chen2011}; Amazonian studies relate fire emissions and probability to drought, forest condition and land use \citep{aragao2018,brando2020,fonseca2019}. These findings motivate antecedent-moisture predictors. The SeasFire cube supports global sub-seasonal to seasonal wildfire modeling at quarter-degree spatial resolution and eight-day temporal resolution \citep{karasante2025}. While \citet{huot2022} benchmark next-day spread conditional on an existing fire, this benchmark predicts VIIRS detections in Bolivia's \num{\GridDegrees}-degree cells during the following \HorizonDays{} days.

FAN's SATRIFO provides monitoring and short-horizon fire-risk information in Bolivia \citep{fanSatrifo}. CIMA documents a Bolivia adaptation of RISICO \citep{cimaBolivia}, and \citet{perello2025} verify the Italian danger-rating system. The reviewed sources do not establish equivalent validation for the Bolivia adaptation. We therefore compare models against climatology and persistence on the same cells and weeks.

\FloatBarrier
\section{Data}
\subsection{Active-fire observations and provenance}
NASA's Fire Information for Resource Management System (FIRMS) supplies the standard-processing active-fire product from the Visible Infrared Imaging Radiometer Suite (VIIRS) on the Suomi National Polar-orbiting Partnership (S-NPP) satellite \citep{schroeder2014,davies2009}. Acquisition spans \AcquisitionStart{} through \AcquisitionEnd{}, including feature warm-up and the final target interval. Primary selection retains presumed vegetation-fire records, with type code \Zero{}, and nominal or high confidence. Confidence classes describe detection quality, not predicted fire probabilities. The all-confidence sensitivity additionally retains low-confidence target observations while keeping primary-selection history predictors unchanged.

Each download is accepted only if its format is valid and its product, area and date range match the recorded request. Checksums and source-row identifiers preserve the download history; overlapping responses are resolved by a fixed rule and exact duplicate records are removed. Acquisition and observation coverage are assessed separately. The first and last acquisition years contain only the dates within this interval.

\par\noindent\begin{minipage}{\linewidth}
\centering\small
\captionsetup{hypcap=false}
{\captionof{table}{Dataset snapshot. Detection and acquisition totals include the warm-up period; cell-weeks span \FirstIssue{} through \LastIssue{}.}\label{tab:dataset}
\begin{tabular}{@{}lS[table-format=8.2]@{}}
\toprule
Quantity & {Value} \\
\midrule
Eligible cells & \EligibleCells \\
Weekly issue times & \IssueWeeks \\
Cell-weeks, including purged rows & \TotalRows \\
National intersection area (\si{\kilo\metre\squared}) & \NationalArea \\
Eligible geographic blocks & \EligibleBlocks \\
Held-out blocks & \HeldOutBlocks \\
Held-out cells & \HeldOutCells \\
Held-out area (\si{\kilo\metre\squared}) & \HeldOutArea \\
Primary national detections & \SelectedDetections \\
All-confidence national detections & \AllConfidenceDetections \\
Acquired days & \AcquisitionDays \\
Unobserved days & \UnobservedDays \\
\bottomrule
\end{tabular}

}
\par\medskip
\includegraphics[width=5.15in]{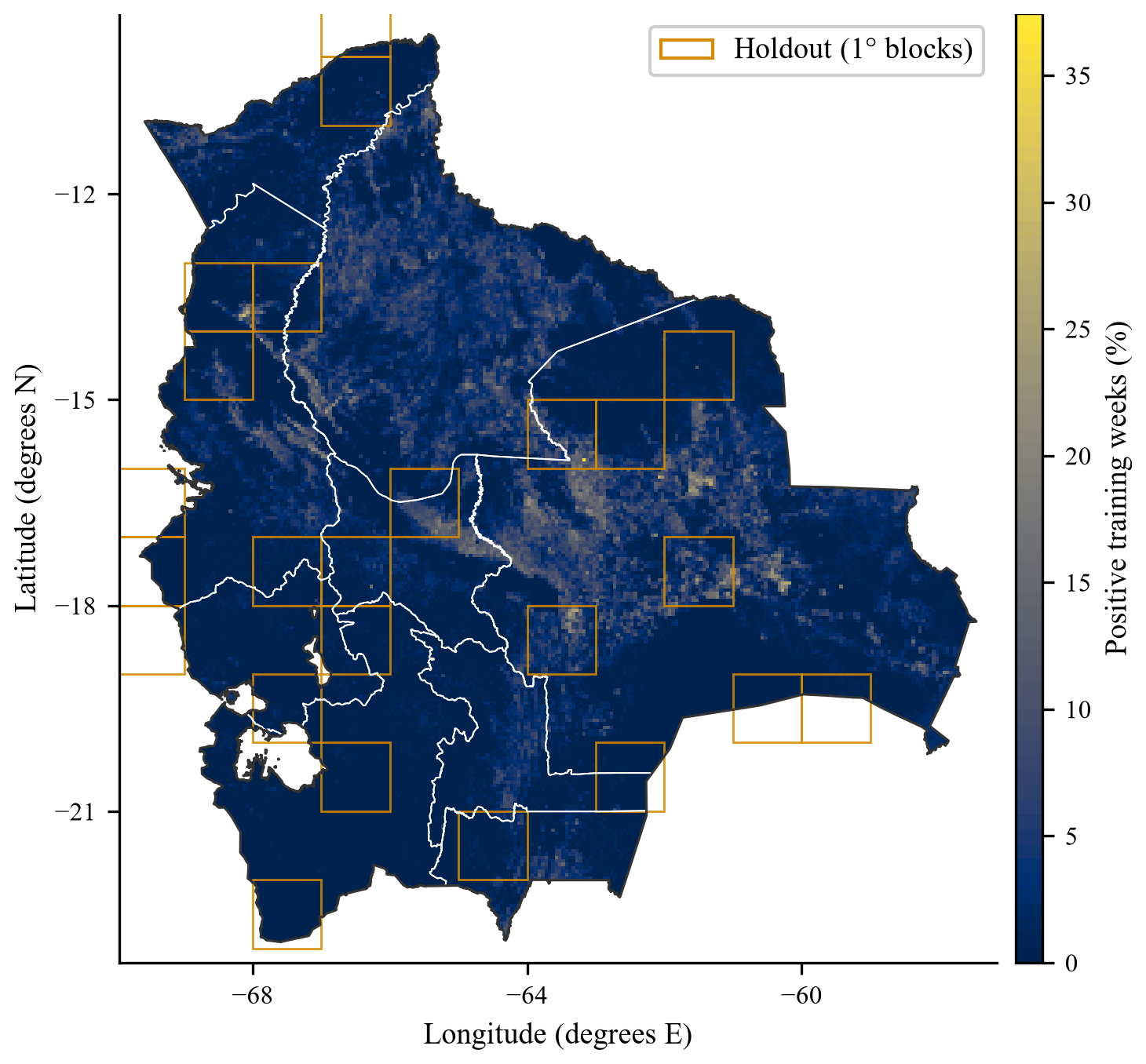}
{\captionof{figure}{Descriptive training-period detection frequency, calculated over complete, non-purged training weeks, and the frozen geographic holdout.}\label{fig:map}}
\end{minipage}\par
\FloatBarrier

\subsection{Boundary and spatial support}
The study uses geoBoundaries gbOpen Bolivia national and departmental polygons \citep{runfola2020} and a global WGS84 grid with \num{\GridDegrees}-degree edges (Figure~\ref{fig:map}). Checksums identify the exact boundary files used. Cells with positive national intersection area are eligible; labels and features include only detections covered by the national polygon. Department assignment uses the largest national intersection, and Santa Cruz is reported separately. Appendix~\ref{app:spatial-details} gives coordinate and boundary conventions.

Inherited interior rings exclude a net \num{\ExcludedArea}\,\si{\kilo\metre\squared} of water and salt bodies after retained islands are accounted for. No additional lake or salt-flat mask is imposed.

Area-budget calculations use the area of each cell inside Bolivia, excluding holes in the national boundary.

\subsection{Observation coverage and outages}\label{sec:coverage}
The NASA MODAPS S-NPP outage table is pinned independently of the acquisition manifest. A UTC day is observed only if it was acquired and the union of documented outage intervals covers strictly less than \OutageHours{} hours. A target with any unobserved day has a missing label, even if some detections are present. An observed zero denotes no selected detection in a fully observed interval.

\begin{figure}[htbp]
\centering
\begin{subfigure}[t]{3.15in}
\includegraphics[width=\linewidth]{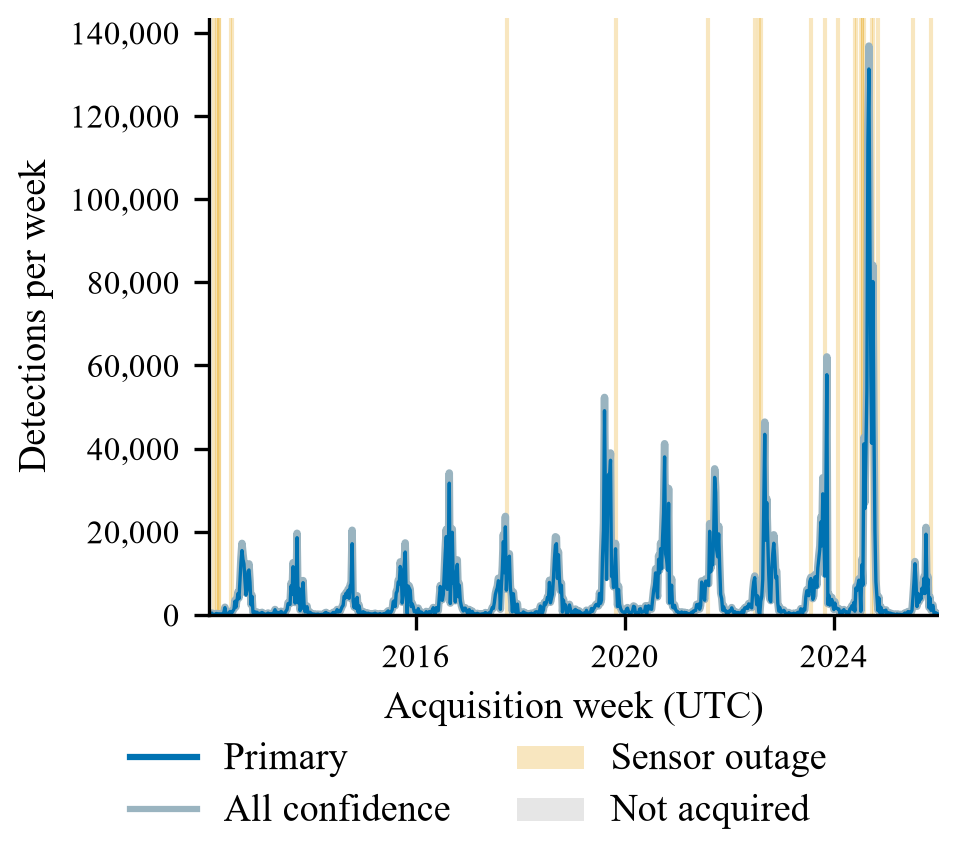}
\caption{Weekly national detections.}\label{fig:weekly}
\end{subfigure}\hfill
\begin{subfigure}[t]{3.15in}
\includegraphics[width=\linewidth]{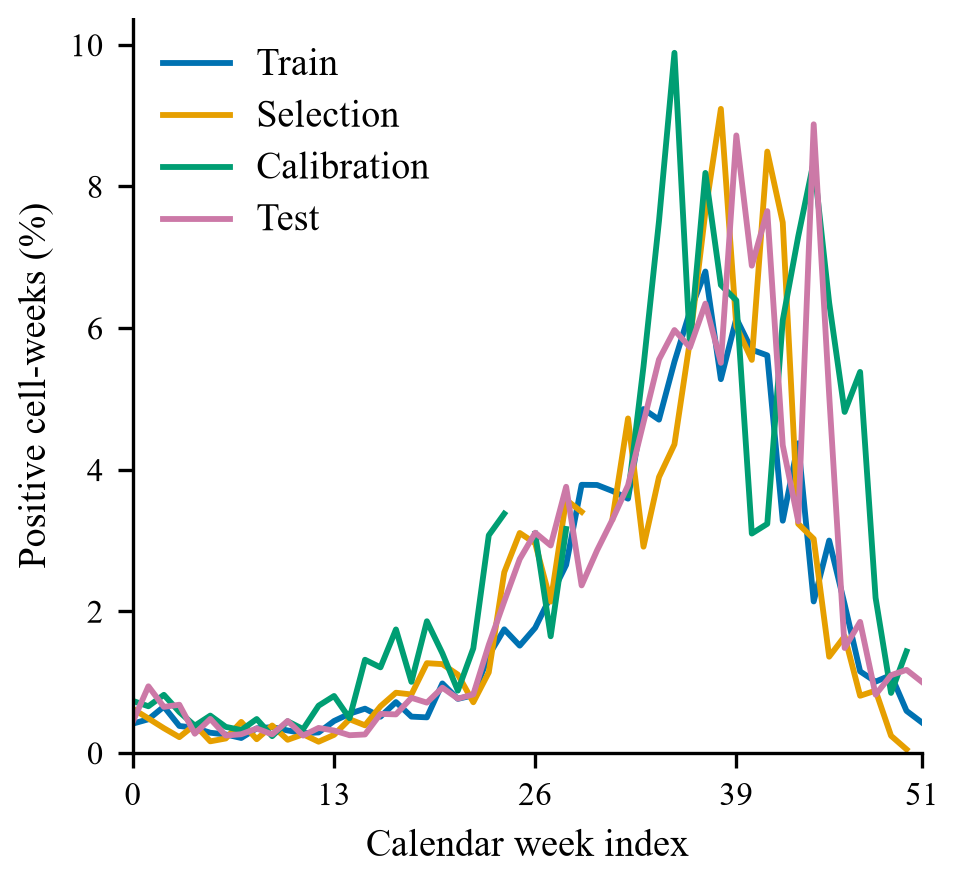}
\caption{Seasonal detection rate.}\label{fig:seasonal}
\end{subfigure}
\caption{Observation coverage and seasonality. Weekly counts retain partial observations during incomplete weeks; shaded intervals mark documented sensor outages. The seasonal cycle uses the frozen calendar-week index and equal weight per labelled cell-week. Both geographic populations enter these descriptive summaries.}
\end{figure}

Fitting and evaluation require complete previous-\HistoryDays{}-day history; incomplete \LongHistoryDays{}-day history remains flagged. A separate sensitivity filter flags fire-season target days with no bounding-box detections. The national series and seasonal cycle are shown in Figure~\ref{fig:weekly}--\ref{fig:seasonal}; Appendix~\ref{app:history} specifies missing-history handling.

\subsection{Retrospective environmental covariates}\label{sec:environmental-data}
The rainfall source is Climate Hazards Center Infrared Precipitation with Stations (CHIRPS) version \ChirpsVersion{} final daily \texttt{rnl}, temporally disaggregated with ERA5 \citep{chirps2025,funk2026}. Its version release date is \ChirpsRelease{}, and final products are normally completed in the following month. Revisions to the historical series prevent a lagged final product from reconstructing its original availability. Each cell receives the value of the CHIRPS pixel containing its center, without interpolation; missing pixel values remain missing. A CHIRPS daily value can combine station observations covering different local reporting periods, so it should not be treated as an identical UTC-day measurement everywhere.

Rainfall predictors summarize earlier totals, dry-day counts and rainfall relative to a training-period reference. Each rainfall comparison uses the same cell-weeks for both models and requires complete rainfall inputs. Appendix~\ref{app:rainfall-features} specifies the windows and reference calculation.

Vegetation predictors use the normalized difference vegetation index (NDVI) and enhanced vegetation index (EVI), derived from quality-screened \ExtensionCompositeDays{}-day MODIS Terra composites (\texttt{\ExtensionProduct{}}) downloaded through AppEEARS \citep{didan2021}. Land-cover predictors use preceding-year \ExtensionLandProduct{} area fractions \citep{mapbiomas2026}. Appendix~\ref{app:covariate-processing} specifies the quality, coverage and age requirements; inputs that fail these requirements remain missing without interpolation.

Vegetation condition and land-cover composition are associated with regional fire occurrence \citep{maillard2020}.

\FloatBarrier
\section{Benchmark definition}
Forecasts are issued at midnight UTC each Monday; $t$ denotes that issue time, and a cell-week is one grid cell during the following \HorizonDays{}-day interval. The target is $y=\One$ if at least one primary detection falls in the national part of an eligible cell during $[t,t+\HorizonDays\text{ days})$, and $y=\Zero$ otherwise, conditional on complete target observation.

\begin{figure}[htbp]
\centering
\includegraphics[width=\linewidth]{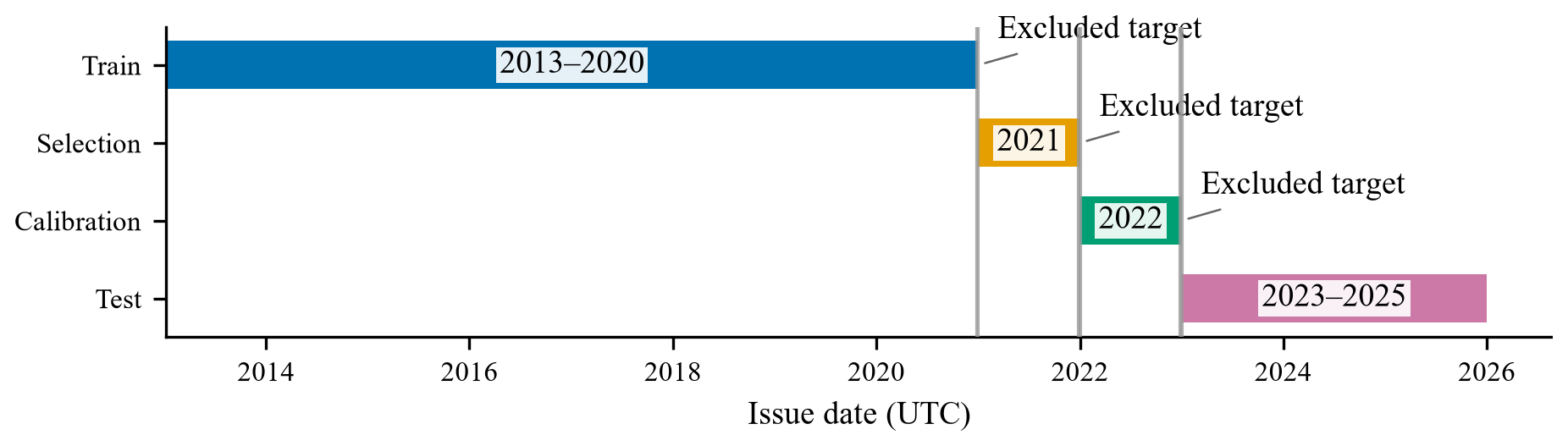}
\caption{Temporal partitions and excluded boundary weeks. Gray marks identify forecasts whose target crosses a partition boundary.}\label{fig:partitions}
\end{figure}

Partitions are assigned by issue date (Figure~\ref{fig:partitions}). We exclude a forecast if its \HorizonDays{}-day target extends into the next partition; a target ending exactly at the boundary is retained. Table~\ref{tab:partitions} distinguishes descriptive row counts from eligible fitting populations.

\begin{table}[htbp]
\centering\small
\caption{Descriptive partition statistics before fitting filters. Labelled share is common to both geographic populations. Positive rates use nonmissing labels; the last column gives the held-out population's positive rate. Test-year counts describe the dataset, not predictive performance.}\label{tab:partitions}
\begin{tabular}{@{}lS[table-format=8.0]S[table-format=3.2]S[table-format=1.2]S[table-format=1.2]@{}}
\toprule
Partition & {Rows} & {Labelled (\%)} & {Positive (\%)} & {Held-out positive (\%)} \\
\midrule
Train & \TrainRows & \TrainLabelPercent & \TrainPositivePercent & \TrainHeldPositivePercent \\
Selection & \SelectRows & \SelectLabelPercent & \SelectPositivePercent & \SelectHeldPositivePercent \\
Calibration & \CalibRows & \CalibLabelPercent & \CalibPositivePercent & \CalibHeldPositivePercent \\
Test & \TestRows & \TestLabelPercent & \TestPositivePercent & \TestHeldPositivePercent \\
Purged & \PurgedRows & \PurgedLabelPercent & \PurgedPositivePercent & \PurgedHeldPositivePercent \\
\bottomrule
\end{tabular}

\end{table}

The geographic test reserves \BlockDegrees{}-degree blocks selected by a fixed hash ranking (Figure~\ref{fig:map}; Table~\ref{tab:dataset}; Appendix~\ref{app:spatial-details}). Held-out labels never enter fitting, selection, calibration or climatology; past observations in those locations may supply antecedent features.

Temporal testing uses non-held-out cells in the test years and the full strict feature set. Geographic testing uses held-out cells in the same years with a separately selected geographic feature set that excludes cell climatology. Both selections use non-held-out validation rows. Consequently, validation scores for the geographic feature set are not geographic-test performance.

\paragraph{Information availability.}\label{sec:availability}
Strict observation-derived predictors use acquisitions before $t$. The frozen leave-one-year-out climatology exception permits other training years, including later years, but excludes validation and test labels. The strict simulation assumes archived standard-processing detections approximate near-real-time inputs available at issue time. Environmental inputs belong to the retrospective variant; their source availability is described in Section~\ref{sec:environmental-data}.

\FloatBarrier
\section{Models and evaluation}
\subsection{Baselines and candidate models}
Climatology estimates the usual detection frequency for a cell and time of year from training data, pooling the corresponding week with the \SmoothingWeeks{} weeks on either side. For the geographic baseline, frequencies are pooled across training cells in the same department because held-out cells cannot supply training labels. Persistence estimates training positive frequencies conditional on previous-\HorizonDays{}-day detection-count bins. Appendix~\ref{app:baseline-details} gives the calendar, clipping and bin definitions.

Strict predictors include cell counts, recency, fire radiative power (FRP), neighborhood and national histories, seasonal sine and cosine, climatology, and a long-history observation flag. Logistic regression \citep{pedregosa2011} uses training-only transformation, imputation and standardization, fitting all positives and a seeded \NegativePercent{}\% negative sample with inverse-sampling weights. LightGBM \citep{ke2017} fits all eligible training rows in \SearchCount{} random-search configurations, with early stopping on selection average precision. The geographic search removes cell climatology. Feature definitions, search spaces and selected settings are in Appendix~\ref{app:specifications}.

Selection ranks uncalibrated AP on \SelectYear{}, breaking ties by model name, to choose the candidate and original baseline before test access. For each model, isotonic calibration \citep{niculescu2005} uses \CalibYear{} data to map its scores monotonically to probabilities that match observed detection frequencies. Primary tables use the original baseline probabilities and calibrated logistic regression and LightGBM probabilities. Calibration-fit diagnostics are reported in Appendix~\ref{app:selection-details}.

\subsection{Metrics, uncertainty and decisions}\label{sec:decisions}
Average precision (AP) summarizes how well the ranking concentrates cell-weeks with detections near its top; higher AP is better. We use the step-wise precision--recall definition, grouping equal scores. Evaluated prevalence accompanies AP because positive outcomes are rare \citep{saito2015}. Brier score and log loss measure probability error; lower values are better. We also report the area under the receiver-operating-characteristic curve (ROC-AUC) and expected calibration error (ECE), the weighted discrepancy between predicted and observed frequencies in \ECEBins{} equally populated probability bins. Log loss clips probabilities at \num{\LogClip} and its complement; undefined subset metrics remain missing with a reason.

Area-budget recall ranks cells by probability within each issue week, breaking ties by stable input order. The budget is \SmallBudget{}\% or \LargeBudget{}\% of eligible national intersection area in the reporting population. If the last cell crosses the area limit, only the fraction needed to reach the limit contributes to recall. Recall is then pooled across weeks. Regional budgets use the region's eligible area.

Uncertainty uses \BootstrapReplicates{} seeded block-bootstrap replicates \citep{kunsch1989}: calendar months for temporal testing and held-out spatial blocks for geographic testing. For each bootstrap draw, both models are evaluated on the same cell-weeks with the same resampled-block weights; rows lacking a finite prediction from either model are excluded. This common set of cell-weeks is the paired support. Percentile intervals omit undefined draws and report valid counts.

H\FirstHypothesis{} requires the \ConfidencePercent{}\% paired month-block interval for calibrated strict AP minus AP of the best original, uncalibrated baseline to lie above \Zero{}. H\SecondHypothesis{} follows the rule stated in the Introduction and additionally requires provenance-verified environmental inputs to be complete on shared support. These are temporal-test decisions; calibrated baselines are a separate sensitivity.

Vegetation and land-cover analyses followed a separate protocol after the primary analysis, with decisions based on validation years and test scores treated as exploratory; the land-cover amendment was specified after vegetation validation results were seen and before the land-cover and combined comparisons accessed fire-detection targets or performance metrics. \ExtensionHThree{}, \ExtensionHFour{} and \ExtensionHFive{} compare vegetation, land cover and both with the saved rainfall model, using independent selection and calibration and identical ordered rows within each model--comparator pair. Their decision requires a finite paired calibrated-AP interval entirely above zero on \CalibYear{} validation data; other or undefined intervals are inconclusive, while unavailable groups are not evaluated. These intervals use the same \CalibYear{} outcomes used to calibrate probabilities, hold the fitted models fixed and are not adjusted for testing multiple hypotheses.

\subsection{Control of test access}\label{sec:test-guard}
Test outcomes were accessed only after primary model selection closed; subsequent evaluations were logged with their reasons.

\FloatBarrier
\section{Results}
\subsection{Temporal generalization and the primary comparison}
Calibrated strict LightGBM improves AP over the validation-selected original climatology baseline by \TAPGain{}, with paired \ConfidencePercent{}\% month-block interval [\HOneLow{}, \HOneHigh{}], supporting H\FirstHypothesis{} (Table~\ref{tab:temporal}). No paired LightGBM--logistic hypothesis was specified.

\begin{table}[htbp]
\centering\footnotesize
\setlength{\tabcolsep}{3pt}
\caption{Primary temporal test on \num{\TBoostRows} eligible non-held-out cell-weeks in \TestYearRange{}. AP intervals use \BootstrapReplicates{} recorded calendar-month-block resamples over \TBlocks{} observed months. R denotes pooled recall of cell-weeks with detections at the stated weekly area budget.}\label{tab:temporal}
\begin{tabular*}{\linewidth}{@{\extracolsep{\fill}}lS[table-format=1.3]rS[table-format=1.3]S[table-format=1.3]S[table-format=1.3]S[table-format=1.4]S[table-format=1.3]S[table-format=1.3]}
\toprule
Model & {AP} & \ConfidencePercent{}\% CI & {ROC-AUC} & {Brier} & {Log loss} & {ECE} & {R@\SmallBudget{}\%} & {R@\LargeBudget{}\%} \\
\midrule
Climatology & \TClimatologyAP & [\num{\TClimatologyAPLow}, \num{\TClimatologyAPHigh}] & \TClimatologyAUC & \TClimatologyBrier & \TClimatologyLogLoss & \TClimatologyECE & \TClimatologyRecallSmall & \TClimatologyRecallLarge \\
Persistence & \TPersistenceAP & [\num{\TPersistenceAPLow}, \num{\TPersistenceAPHigh}] & \TPersistenceAUC & \TPersistenceBrier & \TPersistenceLogLoss & \TPersistenceECE & \TPersistenceRecallSmall & \TPersistenceRecallLarge \\
Logistic regression & \TLogisticAP & [\num{\TLogisticAPLow}, \num{\TLogisticAPHigh}] & \TLogisticAUC & \TLogisticBrier & \TLogisticLogLoss & \TLogisticECE & \TLogisticRecallSmall & \TLogisticRecallLarge \\
LightGBM & \TBoostAP & [\num{\TBoostAPLow}, \num{\TBoostAPHigh}] & \TBoostAUC & \TBoostBrier & \TBoostLogLoss & \TBoostECE & \TBoostRecallSmall & \TBoostRecallLarge \\
\bottomrule
\end{tabular*}

\end{table}

\begin{samepage}
The test contains \num{\ComparisonTestPositives} cell-weeks with detections (\SupplementPrevalence{}\%). Complete-target and recent-history filters retain \SupplementYearAWeeks{}, \SupplementYearBWeeks{} and \SupplementYearCWeeks{} issues in \YearA{}, \YearB{} and \YearC{}: \SupplementRetainedWeeks{} of \SupplementTestWeeks{} test issues. Appendix~\ref{app:exclusions} reports exclusions; Table~\ref{tab:partitions} gives counts before eligibility filters.
\par\end{samepage}

\subsection{Environmental covariates}\label{sec:environmental-results}
The primary \ExtensionHThree{}, \ExtensionHFour{} and \ExtensionHFive{} validation intervals all contain zero, so none meets the decision rule for added information (Table~\ref{tab:environmental-covariates}).

\begin{table}[htbp]
\centering\footnotesize
\setlength{\tabcolsep}{5pt}
\caption{Primary environmental comparisons. H\SecondHypothesis{} contrasts rainfall with strict inputs on the temporal test; the other hypotheses compare additions with the saved rainfall model on non-held-out validation cells. Ordered cell-weeks match within each pair; support differs across hypotheses. The primary lag is \ExtensionLag{} days where applicable. Intervals are \ConfidencePercent{}\% calendar-month-block intervals. AP differences are rounded to four decimals; signed zero preserves the direction of a nonzero endpoint. Decisions use unrounded intervals. For H\SecondHypothesis{}, the saved common-support summary reports an interval but no AP point estimate.}\label{tab:environmental-covariates}
\begin{tabular}{@{}lrS[table-format=7.0]cl@{}}
\toprule
Contrast & Lag (d) & {Paired rows} & 95\% CI for AP difference & Interpretation \\
\midrule
H\SecondHypothesis{} & n/a & \SupplementRainfallRows & $[$\(\HTwoLow\),\,\(\HTwoHigh\)$]$ & Meets frozen H2 rule \\
H3 & \ExtensionLag{} & \ExtensionValidationARows & $[$\(\ExtensionValidationALow\),\,\(\ExtensionValidationAHigh\)$]$ & \ExtensionValidationADecision{} \\
H4 & n/a & \ExtensionValidationBRows & $[$\(\ExtensionValidationBLow\),\,\(\ExtensionValidationBHigh\)$]$ & \ExtensionValidationBDecision{} \\
H5 & \ExtensionLag{} & \ExtensionValidationCRows & $[$\(\ExtensionValidationCLow\),\,\(\ExtensionValidationCHigh\)$]$ & \ExtensionValidationCDecision{} \\
\bottomrule
\end{tabular}

\end{table}

Secondary and exploratory covariate intervals are reported in Appendix~\ref{app:covariate-secondary}; geographic-feature validation uses non-held-out cells. The positive geographic-feature \ExtensionHFour{} interval is secondary. At the \ExtensionLongLag{}-day lag, vegetation alone and the geographic-feature combined model have positive validation intervals; these sensitivity findings do not change the primary decisions. On the reopened temporal test, land-cover and combined intervals are positive while the vegetation interval contains zero; these exploratory scores did not select models or lags. Secondary rainfall metrics are in Appendix~\ref{app:rainfall-secondary}.

\begin{samepage}
\subsection{Geographic transfer}
The geographic LightGBM improves AP over validation-selected original persistence by \GAPGain{}, with paired spatial-block interval [\GAPGainLow{}, \GAPGainHigh{}] (Table~\ref{tab:geographic}).
\par\end{samepage}

\begin{table}[htbp]
\centering\footnotesize
\setlength{\tabcolsep}{3pt}
\caption{Geographic test on \num{\GBoostRows} eligible held-out cell-weeks during \TestYearRange{}. Cell climatology is excluded; the climatology baseline pools departments. AP intervals use \BootstrapReplicates{} resamples of \GBlocks{} held-out spatial blocks. Recall budgets use eligible national intersection area within this population.}\label{tab:geographic}
\begin{tabular*}{\linewidth}{@{\extracolsep{\fill}}lS[table-format=1.3]rS[table-format=1.3]S[table-format=1.3]S[table-format=1.3]S[table-format=1.5]S[table-format=1.3]S[table-format=1.3]}
\toprule
Model & {{AP}} & \ConfidencePercent{}\% CI & {{ROC-AUC}} & {{Brier}} & {{Log loss}} & {{ECE}} & {{R@\SmallBudget{}\%}} & {{R@\LargeBudget{}\%}} \\
\midrule
Climatology & \GClimatologyAP & [\num{\GClimatologyAPLow}, \num{\GClimatologyAPHigh}] & \GClimatologyAUC & \GClimatologyBrier & \GClimatologyLogLoss & \GClimatologyECE & \GClimatologyRecallSmall & \GClimatologyRecallLarge \\
Persistence & \GPersistenceAP & [\num{\GPersistenceAPLow}, \num{\GPersistenceAPHigh}] & \GPersistenceAUC & \GPersistenceBrier & \GPersistenceLogLoss & \GPersistenceECE & \GPersistenceRecallSmall & \GPersistenceRecallLarge \\
Logistic regression & \GLogisticAP & [\num{\GLogisticAPLow}, \num{\GLogisticAPHigh}] & \GLogisticAUC & \GLogisticBrier & \GLogisticLogLoss & \GLogisticECE & \GLogisticRecallSmall & \GLogisticRecallLarge \\
LightGBM & \GBoostAP & [\num{\GBoostAPLow}, \num{\GBoostAPHigh}] & \GBoostAUC & \GBoostBrier & \GBoostLogLoss & \GBoostECE & \GBoostRecallSmall & \GBoostRecallLarge \\
\bottomrule
\end{tabular*}

\end{table}

\subsection{Variation across years and Santa Cruz}
The annual ranking decline in \YearC{} coincides with higher calibration error, while the Santa Cruz subset still favors LightGBM over both baselines (Appendix~\ref{app:subset-scores}).

\subsection{Calibration and area-budget recall}
\begin{figure}[p]
\centering
\begin{subfigure}[t]{\linewidth}
\centering\includegraphics[width=2.58in]{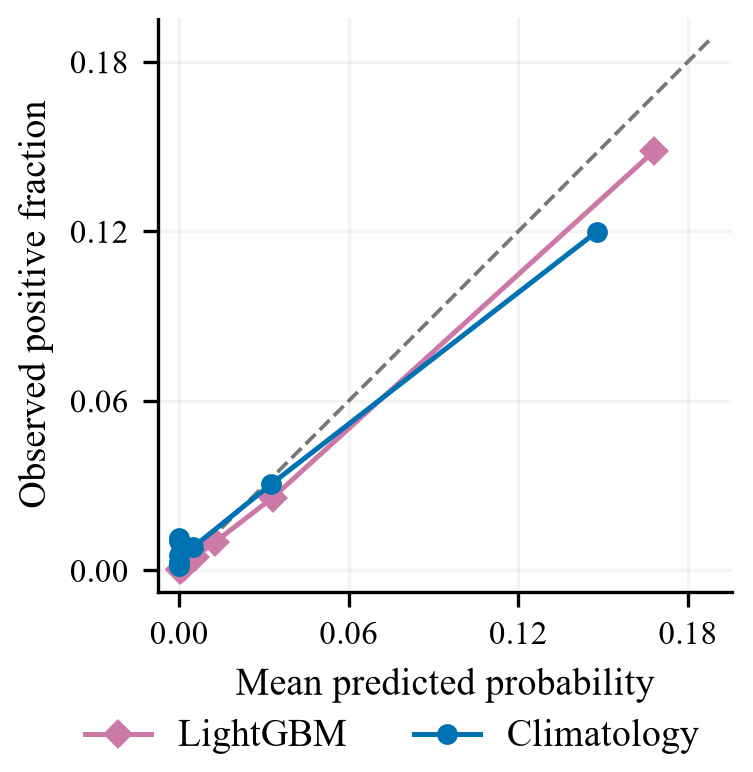}
\caption{Temporal-test calibration.}\label{fig:reliability}
\end{subfigure}
\par\medskip
\begin{subfigure}[t]{\linewidth}
\centering\includegraphics[width=6.15in]{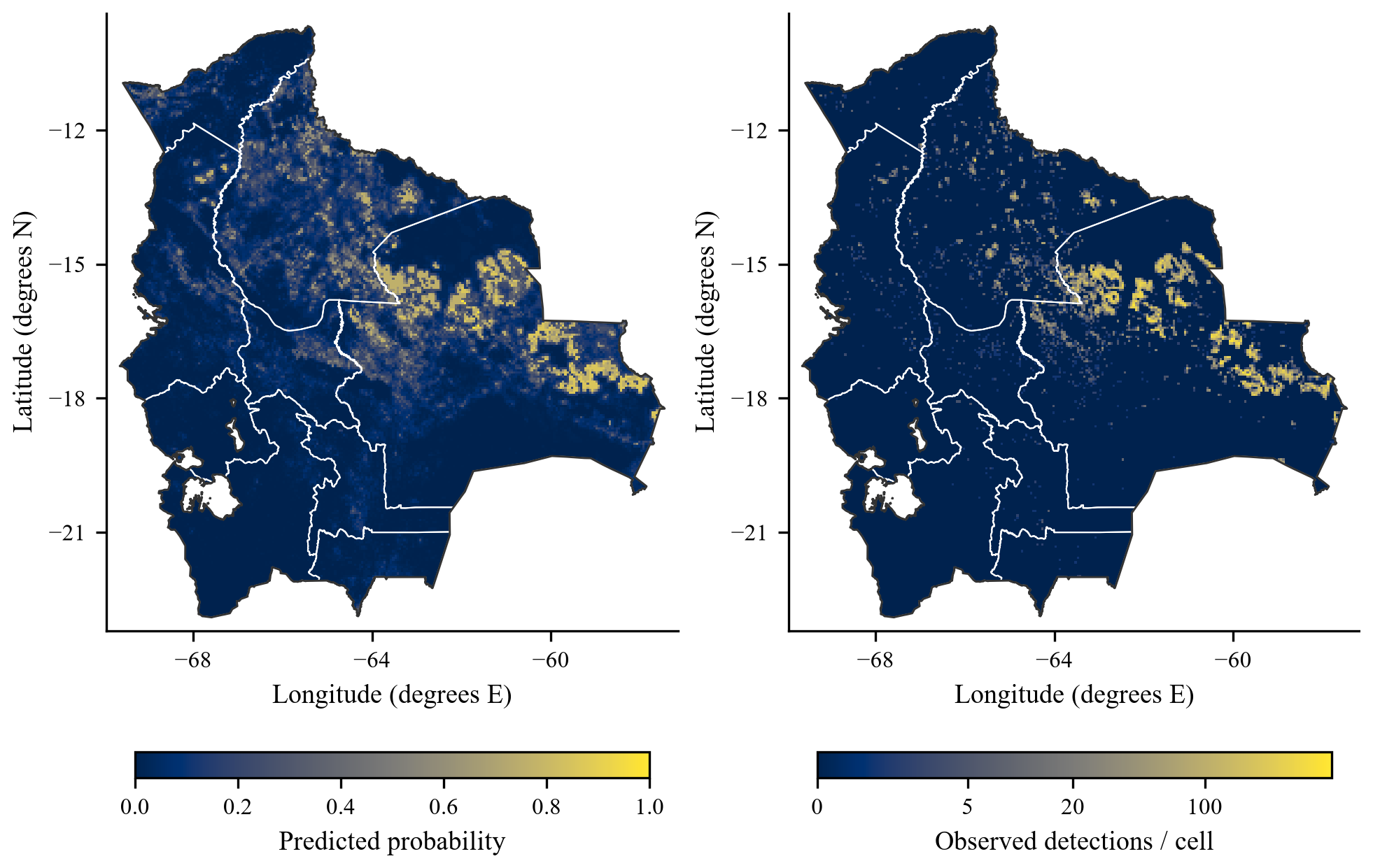}
\caption{Predictions and observed detections.}\label{fig:example}
\end{subfigure}
\caption{Temporal-test calibration in \ECEBins{} equally populated probability bins; the dashed line indicates agreement between predicted probabilities and observed detection frequency. The maps show issue \mbox{\ExampleIssue} and target [\mbox{\ExampleIssue}, \mbox{\ExampleEnd}) UTC, containing \num{\ExampleDetections} detections in \num{\ExamplePositiveCells} cells. The example is the complete \YearB{} week with the largest observed primary detection count among weeks with predictions for all \num{\EligibleCells} cells; ties select the earliest issue. Calibrated strict predictions are used outside held-out blocks and calibrated geographic predictions inside them. Observed cell counts use a $\log(\One+x)$ scale; department outlines use the pinned boundaries.}
\end{figure}

LightGBM reduces aggregate probability error relative to climatology, but its highest-probability bin still overestimates detection frequency (Figure~\ref{fig:reliability}). The higher area-budget recall means that a fixed review area contains more cell-weeks with detections (Tables~\ref{tab:temporal} and~\ref{tab:geographic}).

\subsection{Sensitivity analyses and evaluation support}\label{sec:sensitivities}
Recalibrating the temporal comparator gives a LightGBM-minus-climatology AP interval of \CalibratedBaselineContrast{} (Table~\ref{tab:calibratedbaseline}), retaining a positive contrast under that sensitivity.

\begin{table}[htbp]
\centering\small
\caption{Calibrated-baseline sensitivity on the temporal test. These isotonic baseline probabilities are excluded from the primary comparator definition. The candidate remains the same calibrated strict LightGBM.}\label{tab:calibratedbaseline}
\begin{tabular}{@{}lS[table-format=1.3]S[table-format=1.3]S[table-format=1.4]@{}}
\toprule
Calibrated baseline & {AP} & {Brier} & {ECE} \\
\midrule
Climatology & \CalibratedClimatologyAP & \CalibratedClimatologyBrier & \CalibratedClimatologyECE \\
Persistence & \CalibratedPersistenceAP & \CalibratedPersistenceBrier & \CalibratedPersistenceECE \\
\bottomrule
\end{tabular}

\end{table}

\begin{samepage}
The access log contains \OpeningCount{} openings: the primary evaluation, \SensitivityReportCount{} sensitivity or ablation reopenings (Table~\ref{tab:sensitivities}), one post-hoc learner comparison, and two exploratory covariate evaluations. The complete access log is archived with the dataset \citep{cuellar2026data}.
\par\end{samepage}

\begin{table}[htbp]
\centering\footnotesize
\setlength{\tabcolsep}{4pt}
\renewcommand{\arraystretch}{1.1}
\caption{Sensitivity and ablation reopenings. AP and paired AP differences use the same support within each run. Temporal intervals resample calendar months; geographic intervals resample held-out blocks. Temporal contrasts use original climatology; geographic contrasts use original persistence. All five temporal contrasts meet the H\FirstHypothesis{} rule; H\FirstHypothesis{} is not defined for geographic contrasts. The low-coverage sensitivity excluded no evaluated rows and reproduced the primary scores.}\label{tab:sensitivities}
\begin{tabular}{@{}lS[table-format=1.3]c@{}S[table-format=1.3]@{,\,}S[table-format=1.3]@{}cS[table-format=-1.4]c@{}}
\toprule
Run & {AP} & \multicolumn{4}{c}{95\% AP CI} & {$\Delta$AP} & 95\% difference CI \\
\midrule
\multicolumn{8}{@{}l}{\textit{Temporal}}\\[2pt]
All confidence & \SensitivityATPoint & $[$ & \SensitivityATAPLow & \SensitivityATAPHigh & $]$ & \SensitivityATGainPoint & $[$\(\SensitivityATGainLow\),\,\(\SensitivityATGainHigh\)$]$ \\
Drop climatology & \SensitivityBTPoint & $[$ & \SensitivityBTAPLow & \SensitivityBTAPHigh & $]$ & \SensitivityBTGainPoint & $[$\(\SensitivityBTGainLow\),\,\(\SensitivityBTGainHigh\)$]$ \\
Drop national & \SensitivityCTPoint & $[$ & \SensitivityCTAPLow & \SensitivityCTAPHigh & $]$ & \SensitivityCTGainPoint & $[$\(\SensitivityCTGainLow\),\,\(\SensitivityCTGainHigh\)$]$ \\
Drop neighborhood & \SensitivityDTPoint & $[$ & \SensitivityDTAPLow & \SensitivityDTAPHigh & $]$ & \SensitivityDTGainPoint & $[$\(\SensitivityDTGainLow\),\,\(\SensitivityDTGainHigh\)$]$ \\
\addlinespace[4pt]
\multicolumn{8}{@{}l}{\textit{Geographic}}\\[2pt]
All confidence & \SensitivityAGPoint & $[$ & \SensitivityAGAPLow & \SensitivityAGAPHigh & $]$ & \SensitivityAGGainPoint & $[$\(\SensitivityAGGainLow\),\,\(\SensitivityAGGainHigh\)$]$ \\
Drop climatology & \SensitivityBGPoint & $[$ & \SensitivityBGAPLow & \SensitivityBGAPHigh & $]$ & \SensitivityBGGainPoint & $[$\(\SensitivityBGGainLow\),\,\(\SensitivityBGGainHigh\)$]$ \\
Drop national & \SensitivityCGPoint & $[$ & \SensitivityCGAPLow & \SensitivityCGAPHigh & $]$ & \SensitivityCGGainPoint & $[$\(\SensitivityCGGainLow\),\,\(\SensitivityCGGainHigh\)$]$ \\
Drop neighborhood & \SensitivityDGPoint & $[$ & \SensitivityDGAPLow & \SensitivityDGAPHigh & $]$ & \SensitivityDGGainPoint & $[$\(\SensitivityDGGainLow\),\,\(\SensitivityDGGainHigh\)$]$ \\
\bottomrule
\end{tabular}

\end{table}

The all-confidence run changes target selection, so its AP is not directly comparable with primary-target AP. Removing neighborhood or climatology features lowers temporal AP point estimates; removing national history slightly raises it. These refitted, reselected models lack paired full-versus-ablated intervals and do not identify causal feature contributions or significant between-run differences. Performance on low-coverage rows remains untested.

\FloatBarrier
\section{Discussion and limitations}\label{sec:limitations}
Geographic transfer here retains local observation histories and uses a separately selected model without cell climatology; it does not test locations without historical observations. Different populations, predictors and comparators also preclude interpreting the larger geographic AP as easier generalization.

The inconclusive primary covariate comparisons do not distinguish redundant environmental information from limitations of the chosen features or models. These predictive comparisons do not identify causal effects of rainfall, vegetation condition or land use.

Changing eligible support prevents comparison on identical annual populations, and the descriptive breakdowns do not identify causes. Overprediction in the higher-probability bins of Figure~\ref{fig:reliability} also shows why low pooled ECE cannot guarantee reliability in particular risk groups or future years.

The response measures a satellite detection process. Clouds, smoke, overpass timing, resolution and subpixel intensity affect detection; multiple detections can represent one fire, and observed negatives can contain undetected burning. A single S-NPP sensor supplies no independent cross-sensor confirmation. Scores and area-budget recall therefore cannot be translated into ignition prediction, fractions of fires, burned area or incident severity.

The coverage rules in Section~\ref{sec:coverage} prevent documented outages from becoming negative labels but do not establish uniform viewing conditions. Performance during missing-label intervals is unmeasured. The information assumptions in Section~\ref{sec:availability} likewise remain unverified for near-real-time inputs: archived standard-processing detections provide no overlap with which to test equivalence at issue time.

Reported recall depends on the selected grid and national boundary, including its excluded water and salt-flat areas. The holdout tests one partition without a spatial buffer \citep{roberts2017}. Only \GBlocks{} held-out blocks and \TBlocks{} observed test months contribute to the primary intervals. Fire activity can remain correlated across neighboring months or blocks; because block size was not varied, the intervals' nominal coverage was not checked under alternative dependence scales. Intervals condition on fixed fits and partitions; they omit model-selection uncertainty, sensor error and boundary choices.

\FloatBarrier
\section{Conclusions}
Earlier satellite detections improved next-week rankings in the evaluated Bolivian cells. On the \TestYearRange{} temporal test, calibrated LightGBM achieved average precision (AP) of \TBoostAP{} against \TClimatologyAP{} for the selected climatology baseline; the paired \ConfidencePercent{}\% interval for the AP gain was \HOneLow{}--\HOneHigh{}. A separately selected geographic model achieved AP \GBoostAP{} against \GPersistenceAP{} for persistence in regions excluded from fitting. Those regions still supplied their own past observations, so this comparison concerns transfer to places with a detection history. Performance where such histories are unavailable remains untested.

At a weekly review budget covering \LargeBudget{}\% of eligible area in the temporal-test population, the model would have covered \TBoostRecallLargePercent{}\% of cell-weeks with detections, compared with \TClimatologyRecallLargePercent{}\% for climatology. This result supports evaluating ranked cells as a way to direct satellite review and field verification. It does not establish an allocation rule for response crews. Scores measure satellite detections, not individual fires or their causes, and probability estimates still overpredict in the highest-probability group. Coverage filters retained \SupplementRetainedWeeks{} of \SupplementTestWeeks{} test issues, so the reported performance does not describe all weeks.

Rainfall produced no clear AP improvement. The primary validation comparisons for vegetation, land cover and combined inputs were inconclusive; positive secondary and exploratory intervals cannot change those decisions. These results leave open whether different environmental features would improve prediction.

The next step is prospective evaluation during the burning season in departments with high monitoring demand. Each weekly input snapshot and prediction should be archived before its target week, then assessed for ranking, calibration and usefulness at the agency's review capacity. Alternative learners need evaluation on previously unobserved weeks before any model change. Adding NOAA-20/21 observations also requires cross-sensor consistency and observation-completeness checks.

\FloatBarrier
\section{Declarations and availability}
\paragraph{Funding.}
This research received no external funding.

\paragraph{Competing interests.}
The author declares no competing interests.

\paragraph{Author contribution.}
Victor Hugo Cuellar Flores performed the conceptualization, methodology, software development, analysis and writing.

\paragraph{Ethics.}
This analysis of open remote-sensing and geographic data involved no human participants or animal experiments.

\paragraph{Acknowledgements.}
The author acknowledges NASA FIRMS, LANCE and ESDIS; NASA LP DAAC and AppEEARS; the Climate Hazards Center for CHIRPS; MapBiomas Bolivia; geoBoundaries; and GeoNames for the source data and distribution services.

\paragraph{Code and data availability.}
Version \texttt{0.1.0} of the \emph{Bolivia Fire benchmark data and reproduction artifacts} dataset is archived on Zenodo \citep{cuellar2026data}. It contains the benchmark panel and covariates, frozen configurations, the environment lockfile, fitted models and calibrators, saved predictions, evaluation and access records, and permitted raw-source snapshots. Compiled and derived data are licensed under CC BY \CCVersion{}; raw sources retain their upstream terms. Global CHIRPS rasters, MODIS HDF granules, native MapBiomas maps and operational live forecasts are excluded. The MIT-licensed software is available at \url{https://github.com/Cupesi/bolivia-fire}, release \texttt{v0.1.0} \citep{cuellar2026software}. The dataset records each experiment's producing code revision; primary training used the \href{https://github.com/Cupesi/bolivia-fire/tree/\SupplementCodeCommit}{recorded immutable revision}.

\paragraph{Generative-AI use.}
Generative AI tools were used to assist with software development, data processing and analysis; the author reviewed the outputs and takes responsibility for the work.

\begingroup\small
\IdentifierBreaks\raggedright
\interlinepenalty=10000
\setlength{\bibsep}{2pt plus 1pt minus 1pt}
\Needspace{4\baselineskip}
\bibliographystyle{plainnat}
\bibliography{references}
\endgroup
\appendix
\section{Post-hoc comparison of alternative learners}\label{app:alternative-learners}
This comparison was designed and run after the primary test opening, outside the frozen protocol; all of its results are exploratory. It assesses sensitivity to learner choice without changing the primary hypotheses, comparator or reported model.

XGBoost hist \citep{chen2016} tests a different histogram-tree implementation with depth-wise growth. CatBoost \citep{prokhorenkova2018} tests symmetric trees and regularization without adding categorical predictors. A spatial convolutional neural network (CNN) tests learned directional local interactions on weekly national grids. The learners use identical eligible target rows, strict feature types and metric definitions, including boundary purges, geographic-label exclusion and the complete recent-history filter. The CNN uses neighboring cells' predictors as additional spatial context, so its input representation differs from the tabular learners' row-wise representation.

Appendix~\ref{app:specifications} gives the search budgets, training and selected settings. Alternatives use the same selection/calibration partition roles as the primary learners, with trial-index tie resolution and no validation refitting.

\begin{table}[htbp]
\centering\small
\caption{Post-hoc validation comparison against the saved strict LightGBM comparator in Table~\ref{tab:validation}. Selection AP is uncalibrated; calibration AP and the paired alternative-minus-LightGBM difference use calibrated scores on \texttt{val\_calib}. Intervals are \ConfidencePercent{}\% paired calendar-month percentile intervals.}\label{tab:alternative-validation}
\begin{tabular}{@{}lS[table-format=1.3]S[table-format=1.3]S[table-format=-1.4]c@{}S[table-format=-1.4]@{,\,}S[table-format=1.4]@{}c@{}}
\toprule
Learner & {Select AP} & {Calib AP} & \multicolumn{5}{c}{$\Delta$ AP [\ConfidencePercent\% CI]} \\
\midrule
XGBoost hist & \ComparisonXGBoostSelectAP & \ComparisonXGBoostCalibAP & \ComparisonXGBoostGain & $[$ & \ComparisonXGBoostGainLow & \ComparisonXGBoostGainHigh & $]$ \\
CatBoost & \ComparisonCatBoostSelectAP & \ComparisonCatBoostCalibAP & \ComparisonCatBoostGain & $[$ & \ComparisonCatBoostGainLow & \ComparisonCatBoostGainHigh & $]$ \\
Spatial CNN & \ComparisonCNNSelectAP & \ComparisonCNNCalibAP & \ComparisonCNNGain & $[$ & \ComparisonCNNGainLow & \ComparisonCNNGainHigh & $]$ \\
\bottomrule
\end{tabular}

\end{table}
\FloatBarrier

Table~\ref{tab:alternative-validation} uses \ComparisonBootstrapReplicates{} paired month-block resamples with seed \ComparisonSeed{} and identical rows and month multiplicities within each comparison. Only \ComparisonCalibrationBlocks{} observed months contribute to the calibration-year intervals. Because these intervals reuse the calibration outcomes and hold fitted models fixed, they are not an independent assessment of calibrated probabilities. None establishes an AP gain over LightGBM. Isotonic ties can reduce AP.

\FloatBarrier
\paragraph{CatBoost reproducibility.}
GPU floating-point reduction nondeterminism is consistent with the observed score differences despite identical training/selection arrays, trial parameters and CatBoost binary. The selected CatBoost trial changed from index \ComparisonCatSourceTrial{} to \ComparisonCatRerunTrial{} between recorded runs, although neither validation interval established superiority over LightGBM. The earlier run is recorded in \path{artifacts/evaluation/comparison_reference.json}. A separate CPU audit fitted the GPU-selected configuration at its fixed \ComparisonCatCPUTrees{}-tree count, obtaining selection AP \ComparisonCatCPUSelectionAP{} and an exact saved-model prediction roundtrip. This single refit does not establish between-fit determinism; it was not calibrated, scored on test or used for model selection.

\subsection{Exploratory test results}
Table~\ref{tab:alternative-test} uses the same eligible non-held-out test population as Table~\ref{tab:temporal}; it does not evaluate geographic holdout transfer. Paired test intervals were not recorded.

\begin{table}[htbp]
\centering\small
\caption{Exploratory test AP in \TestYearRange{}. The primary calibrated LightGBM score is in Table~\ref{tab:temporal}; its uncalibrated score is shown here.}\label{tab:alternative-test}
\begin{tabular}{@{}lS[table-format=1.3]S[table-format=1.3]@{}}
\toprule
Learner & {Raw AP} & {Calibrated AP} \\
\midrule
LightGBM (recorded) & \ComparisonLightGBMTestRawAP & \multicolumn{1}{c}{---} \\
XGBoost hist & \ComparisonXGBoostTestRawAP & \ComparisonXGBoostTestCalibratedAP \\
CatBoost & \ComparisonCatBoostTestRawAP & \ComparisonCatBoostTestCalibratedAP \\
Spatial CNN & \ComparisonCNNTestRawAP & \ComparisonCNNTestCalibratedAP \\
\bottomrule
\end{tabular}

\end{table}

\section{Feature and learner specifications}\label{app:specifications}
\subsection{Baseline and calendar definitions}\label{app:baseline-details}
The frozen calendar-week index and seasonal angle are
\[
 w(t)=\min\!\left(\left\lfloor\frac{\operatorname{dayofyear}(t)-\One}{\HorizonDays}\right\rfloor,\LastWeekIndex\right),
 \qquad \theta(t)=\Two\pi w(t)/\CalendarWeeks.
\]
Climatology smooths circularly across the calendar-year boundary. Department climatology weights non-held-out cell-weeks equally. Training rows exclude their own calendar year's labels; the temporal baseline clips cell climatology to $[\num{\LowerClip},\One-\num{\LowerClip}]$. Persistence bins the previous-\HorizonDays{}-day count as $\Zero$, $\One$, $\Two$--$\SupplementPersistenceUpper$ and $\geq\SupplementPersistenceHigh$, using overall training frequency for an empty bin.

\subsection{Spatial and rainfall processing}\label{app:spatial-details}
National membership includes boundary points. Coordinate snapping to \num{\CoordinateSnap}\,degrees precedes half-open cell assignment. Department ties use deterministic name ordering. Interior-ring exclusions are recorded in \path{docs/dataset.md}; they do not quantify water clipped by an exterior boundary. Cell areas use WGS84 geodesic intersections. Geographic blocks are ranked by SHA256 of seed \HoldoutSeed{} and block identifier; the lowest \HoldoutPercent{}\% are selected after flooring the block count, without imposing equal area. Outage intervals are half-open, clipped at midnight and unioned before daily duration is calculated.

\paragraph{Rainfall features.}\label{app:rainfall-features}
Antecedent sums cover \RainShortDays{}, \RainMiddleDays{} and \RainLongDays{} days. Features also count days below \DryThreshold{}\,mm over the middle window and divide long-window rainfall by the cell's training-year reference mean. Windows exclude the issue day. Training references exclude the issue year; later issues use all training reference years. Missing or zero reference means remain missing, and completeness flags gate eligible rows.

\subsection{Strict predictors and missingness}\label{app:history}
Table~\ref{tab:features} lists the \ParameterAFeatureCount{} strict predictors in their recorded order. The geographic model removes only \mbox{\path{train_climatology_cell}}, leaving \ParameterBFeatureCount{} predictors. Definitions follow \path{src/bolivia_fire/features.py} and the frozen \path{configs/experiments.yaml}.

\begingroup\small
\begin{longtable}{@{}>{\raggedright\arraybackslash}p{0.34\linewidth}>{\raggedright\arraybackslash}p{\dimexpr0.66\linewidth-2\tabcolsep\relax}@{}}
\caption{Exact strict predictors, definitions and windows. All observation windows exclude issue time $t$.}\label{tab:features}\\
\toprule Feature name & Definition, window and unit \\\midrule\endfirsthead
\multicolumn{2}{l}{\tablename\ \ref{tab:features}\ continued}\\\toprule Feature name & Definition, window and unit \\\midrule\endhead
\bottomrule\endfoot
\texttt{\SupplementFeatureAName{}} & Selected in-country cell detections in $[t-\SupplementDaysShort,t)$ days; count. \\[3pt]
\texttt{\SupplementFeatureBName{}} & Selected in-country cell detections in $[t-\SupplementDaysFortnight,t)$ days; count. \\[3pt]
\texttt{\SupplementFeatureCName{}} & Selected in-country cell detections in $[t-\SupplementDaysMonth,t)$ days; count. \\[3pt]
\texttt{\SupplementFeatureDName{}} & Selected in-country cell detections in $[t-\SupplementDaysQuarter,t)$ days; count. \\[3pt]
\texttt{\SupplementFeatureEName{}} & Selected in-country cell detections in $[t-\SupplementDaysYear,t)$ days; count. \\[3pt]
\texttt{\SupplementFeatureFName{}} & Indicator that the previous-\SupplementDaysShort{}-day cell count exceeds \Zero; binary. \\[3pt]
\texttt{\SupplementFeatureGName{}} & Days since the latest prior selected cell detection, capped at \SupplementRecencyCap; the cap is also used if no last date is retained. \\[3pt]
\texttt{\SupplementFeatureHName{}} & No known selected cell detection before $t$, including the older-history summary; binary. \\[3pt]
\texttt{\SupplementFeatureIName{}} & Sum of reported fire radiative power over selected cell detections in $[t-\SupplementDaysMonth,t)$ days; MW, summed retrievals rather than energy. \\[3pt]
\texttt{\SupplementFeatureJName{}} & Selected cell detection count in $[t-\SupplementPreviousStart,t-\SupplementPreviousEnd)$ days; a fixed day offset rather than a calendar-year join. \\[3pt]
\texttt{\SupplementFeatureKName{}} & Selected detection count over the $\SupplementNeighborhoodSmall\times\SupplementNeighborhoodSmall$ grid neighborhood, including the center, in $[t-\SupplementDaysShort,t)$ days. \\[3pt]
\texttt{\SupplementFeatureLName{}} & Selected detection count over the $\SupplementNeighborhoodLarge\times\SupplementNeighborhoodLarge$ grid neighborhood, including the center, in $[t-\SupplementDaysShort,t)$ days. \\[3pt]
\texttt{\SupplementFeatureMName{}} & Selected detection count over the $\SupplementNeighborhoodSmall\times\SupplementNeighborhoodSmall$ grid neighborhood, including the center, in $[t-\SupplementDaysMonth,t)$ days. \\[3pt]
\texttt{\SupplementFeatureNName{}} & Selected detection count over the $\SupplementNeighborhoodLarge\times\SupplementNeighborhoodLarge$ grid neighborhood, including the center, in $[t-\SupplementDaysMonth,t)$ days. \\[3pt]
\texttt{\SupplementFeatureOName{}} & Selected detections across the national polygon in $[t-\SupplementDaysShort,t)$ days; count. \\[3pt]
\texttt{\SupplementFeaturePName{}} & Selected detections across the national polygon in $[t-\SupplementDaysMonth,t)$ days; count. \\[3pt]
\texttt{\SupplementFeatureQName{}} & $\sin\theta(t)$ from the frozen week index; dimensionless, current issue calendar. \\[3pt]
\texttt{\SupplementFeatureRName{}} & $\cos\theta(t)$ from the frozen week index; dimensionless, current issue calendar. \\[3pt]
\texttt{\SupplementFeatureSName{}} & Cell positive-cell-week frequency in training years, circularly smoothed by $\pm\SmoothingWeeks$ week indices; probability. \\[3pt]
\texttt{\SupplementFeatureTName{}} & Department-pooled training positive frequency with equal weight per eligible non-held-out cell-week and the same smoothing; probability. \\[3pt]
\texttt{\SupplementFeatureUName{}} & Every day in $[t-\SupplementDaysYear,t)$ acquired and observed under the outage rule; binary. \\[3pt]
\end{longtable}

\endgroup

Counts and FRP sums retain the available selected observations when a longer history is incomplete; an empty aggregation is zero, without implying complete observation. Missing climatology remains missing when its training denominator is empty. Neighborhoods contain only national cells.

\subsection{Selected primary learners and calibrators}
Selected LightGBM parameters and retained tree counts are recorded in the strict and geographic \texttt{run.json} training records.

\begin{table}[htbp]
\centering\footnotesize
\caption{Selected primary LightGBM configurations; fitting uses all eligible training rows.}\label{tab:primary-settings}
\begin{tabular}{@{}lS[table-format=4.2]S[table-format=3.2]@{}}
\toprule
Setting & {Temporal} & {Geographic} \\
\midrule
Leaves & \SupplementTemporalNumLeaves & \SupplementGeographicNumLeaves \\
Learning rate & \SupplementTemporalLearningRate & \SupplementGeographicLearningRate \\
Minimum leaf rows & \SupplementTemporalMinDataInLeaf & \SupplementGeographicMinDataInLeaf \\
Feature fraction & \SupplementTemporalFeatureFraction & \SupplementGeographicFeatureFraction \\
L2 penalty & \SupplementTemporalLambdaL & \SupplementGeographicLambdaL \\
Selected tree count & \SupplementTemporalBestIteration & \SupplementGeographicBestIteration \\
\bottomrule
\end{tabular}

\end{table}
\FloatBarrier

The primary search uses seed \SearchSeed{} and the frozen leaf counts $\{\SearchLeaves\}$, learning rates $\{\SearchRates\}$, minimum leaf sizes $\{\SearchMinimumLeaves\}$, feature fractions $\{\SearchFractions\}$ and quadratic penalties $\{\SearchPenalties\}$. Common settings are binary objective, AP stopping metric, deterministic column-wise histograms, \HistogramBins{} bins and \ModelThreads{} threads. The maximum \BoostRounds{} rounds and \EarlyStopRounds{}-round patience are search limits; the selected tree counts in Table~\ref{tab:primary-settings} are the retained iterations.

\begin{samepage}
Logistic regression uses scikit-learn \SupplementSklearnVersion{}, solver \texttt{\SupplementLogisticSolver{}}, maximum \SupplementLogisticIterations{} iterations and seed \SearchSeed{}; remaining settings are that recorded version's constructor defaults. This is a fixed configuration rather than a logistic hyperparameter search. Both feature sets fit all positives and negative fraction \SupplementLogisticFraction{}; the temporal fit contains \num{\SupplementLogisticSampledRows} sampled rows with negative weight \SupplementLogisticWeight{} and positive weight \One{}. Transform means and scales use the complete eligible training population before negative sampling. Wholly missing transformed training columns use mean \Zero{}; negligible-variance columns use scale \One{}. Signed seasonal terms and binary flags bypass $\log(\One+x)$, then every column is mean-imputed and standardized. LightGBM handles missing predictors natively without this transform.
\par\end{samepage}

The selected LightGBM and logistic models use a monotone isotonic regression fitted only on \CalibYear{}, bounded to $[\Zero,\One]$, with out-of-range inputs clipped. The same calibration procedure is applied separately to the alternative learners after their selections close. Unavailable or single-class calibration retains the identity map and records the reason. The recorded selection files identify the exact primary model and calibrator SHA256 checksums.

\subsection{Alternative-learner search and selected settings}
The ordered \ComparisonConfigurations{} draws per family, search spaces and seed \ComparisonSeed{} are recorded in \path{configs/comparison.yaml}; Table~\ref{tab:alternative-spaces} lists the spaces. Selection uses full \SelectYear{} uncalibrated exact AP and ties keep the first trial. Selected configurations are recorded in \path{artifacts/evaluation/comparison_selection.json}; trial indices are zero-based.

\begin{table}[htbp]
\captionsetup{font=small}
\begin{WideTablePanel}
\centering\footnotesize
\AlignedPanelCaption{\caption{Frozen alternative-learner search spaces. Ordered sampled configurations are retained in the configuration snapshot.}\label{tab:alternative-spaces}}
\begin{tabular}{@{}lll@{}}
\toprule
Learner & Search parameter & Frozen choices \\
\midrule
XGBoost & max\_depth & \SupplementSearchXgboostMaxDepth{} \\
XGBoost & eta & \SupplementSearchXgboostEta{} \\
XGBoost & min\_child\_weight & \SupplementSearchXgboostMinChildWeight{} \\
XGBoost & colsample\_bytree & \SupplementSearchXgboostColsampleBytree{} \\
XGBoost & lambda & \SupplementSearchXgboostLambda{} \\
CatBoost & depth & \SupplementSearchCatboostDepth{} \\
CatBoost & learning\_rate & \SupplementSearchCatboostLearningRate{} \\
CatBoost & l2\_leaf\_reg & \SupplementSearchCatboostLLeafReg{} \\
CatBoost & random\_strength & \SupplementSearchCatboostRandomStrength{} \\
CatBoost & bagging\_temperature & \SupplementSearchCatboostBaggingTemperature{} \\
CNN & width & \SupplementSearchSpatialcnnWidth{} \\
CNN & depth & \SupplementSearchSpatialcnnDepth{} \\
CNN & learning\_rate & \SupplementSearchSpatialcnnLearningRate{} \\
CNN & weight\_decay & \SupplementSearchSpatialcnnWeightDecay{} \\
CNN & dropout & \SupplementSearchSpatialcnnDropout{} \\
\bottomrule
\end{tabular}

\end{WideTablePanel}\hfill
\begin{NarrowTablePanel}
\centering\footnotesize
\AlignedPanelCaption{\caption{Selected XGBoost histogram configuration, trial \SupplementXGBTrial{}. The default evaluation metric is disabled; exact selection AP controls \EarlyStopRounds{}-round patience within \BoostRounds{} rounds.}\label{tab:xgb-settings}}
\begin{tabular}{@{}lS[table-format=2.2]@{}}
\toprule
Setting & {Selected value} \\
\midrule
tree\_method & \multicolumn{1}{c}{\SupplementXGBTreeMethod{}} \\
device & \multicolumn{1}{c}{\SupplementXGBDevice{}} \\
max\_bin & \SupplementXGBMaxBin \\
objective & \multicolumn{1}{c}{\SupplementXGBObjective{}} \\
seed & \SupplementXGBSeed \\
nthread & \SupplementXGBNthread \\
subsample & \SupplementXGBSubsample \\
max\_depth & \SupplementXGBMaxDepth \\
eta & \SupplementXGBEta \\
min\_child\_weight & \SupplementXGBMinChildWeight \\
colsample\_bytree & \SupplementXGBColsampleBytree \\
lambda & \SupplementXGBLambda \\
Selected tree count & \SupplementXGBBest \\
\bottomrule
\end{tabular}

\end{NarrowTablePanel}
\end{table}

\begin{table}[htbp]
\captionsetup{font=small}
\begin{WideTablePanel}
\centering\footnotesize
\AlignedPanelCaption{\caption{Selected CatBoost GPU configuration, trial \SupplementCatTrial{}. Each trial fits the full iteration budget; exact AP at every \SupplementCatCheckpoint{} trees chooses the retained prefix, without a separate early-stopping fit.}\label{tab:cat-settings}}
\begin{tabular}{@{}lS[table-format=3.2]@{}}
\toprule
Setting & {Selected value} \\
\midrule
task\_type & \multicolumn{1}{c}{\SupplementCatTaskType{}} \\
iterations & \SupplementCatIterations \\
loss\_function & \multicolumn{1}{c}{\SupplementCatLossFunction{}} \\
border\_count & \SupplementCatBorderCount \\
random\_seed & \SupplementCatRandomSeed \\
thread\_count & \SupplementCatThreadCount \\
bootstrap\_type & \multicolumn{1}{c}{\SupplementCatBootstrapType{}} \\
depth & \SupplementCatDepth \\
learning\_rate & \SupplementCatLearningRate \\
l2\_leaf\_reg & \SupplementCatLLeafReg \\
random\_strength & \SupplementCatRandomStrength \\
bagging\_temperature & \SupplementCatBaggingTemperature \\
Selected tree count & \SupplementCatBest \\
\bottomrule
\end{tabular}

\end{WideTablePanel}\hfill
\begin{NarrowTablePanel}
\centering\footnotesize
\AlignedPanelCaption{\caption{Selected spatial CNN configuration, trial \SupplementCNNTrial{}. Depth counts the spatial convolutional layers after the stem.}\label{tab:cnn-settings}}
\begin{tabular}{@{}lS[table-format=2.4]@{}}
\toprule
Setting & {Selected value} \\
\midrule
width & \SupplementCNNWidth \\
depth & \SupplementCNNDepth \\
learning\_rate & \SupplementCNNLearningRate \\
weight\_decay & \SupplementCNNWeightDecay \\
dropout & \SupplementCNNDropout \\
Selected epoch & \SupplementCNNBest \\
\bottomrule
\end{tabular}

\end{NarrowTablePanel}
\end{table}

\begin{samepage}
The CNN maps \ParameterAFeatureCount{} strict channels onto weekly national grids, using the same training-only log transform, imputation and standardization as logistic regression. Inputs outside national cells are zero after normalization. Held-out locations supply past-feature context, while their labels are excluded from normalization, fitting, selection and calibration. Other ineligible target cells are masked from the loss.

The CNN has a $\SupplementPointKernel\times\SupplementPointKernel$ convolutional stem, the selected number of $\SupplementSpatialKernel\times\SupplementSpatialKernel$ spatial layers, and a $\SupplementPointKernel\times\SupplementPointKernel$ output layer with a linear skip from the center cell. Hidden layers use Gaussian error linear unit (GELU) activations \citep{hendrycks2016}; spatial layers use same-size padding and dropout. The head and skip weights start at zero; head bias is the logit of training prevalence. Training uses every eligible row once per epoch in unweighted binary cross-entropy with logits, with no negative sampling. AdamW \citep{loshchilov2019} uses the selected learning rate and weight decay, gradient-norm clipping at \SupplementGradientClip{}, \SupplementCNNBatchWeeks{} shuffled weeks per batch, at most \SupplementCNNEpochs{} epochs and \SupplementCNNPatience{} epochs of selection-AP patience.
\par\end{samepage}

CUDA deterministic algorithms are enabled, cuDNN benchmarking is disabled and mixed precision is disabled. The selected checkpoint uses the epoch shown in Table~\ref{tab:cnn-settings}; learners are never refitted on validation rows.

\subsection{Selected models}\label{app:selection-details}

\begin{table}[htbp]
\centering\footnotesize
\setlength{\tabcolsep}{3pt}
\caption{Selected LightGBM identifiers, predictor counts and comparators from tracked closed selection records. Both rainfall variants add only the frozen antecedent CHIRPS features.}\label{tab:parameters}
\begin{tabular}{@{}llS[table-format=2.0]l@{}}
\toprule
Feature set & Selected model & {Predictors} & Comparator \\
\midrule
Strict temporal & \texttt{\ParameterAModel{}} & \ParameterAFeatureCount & \ParameterABaseline{} \\
Strict geographic & \texttt{\ParameterBModel{}} & \ParameterBFeatureCount & \ParameterBBaseline{} \\
Rainfall temporal & \texttt{\ParameterCModel{}} & \ParameterCFeatureCount & \ParameterCBaseline{} \\
Rainfall geographic & \texttt{\ParameterDModel{}} & \ParameterDFeatureCount & \ParameterDBaseline{} \\
\bottomrule
\end{tabular}

\end{table}

\ValidationTableSpace
\begingroup\footnotesize
\setlength{\tabcolsep}{4pt}
\begin{longtable}{@{}lS[table-format=1.3]S[table-format=1.3]S[table-format=1.3]S[table-format=1.6]@{}}
\caption{Validation diagnostics. Select AP uses uncalibrated \SelectYear{} predictions; Calib columns use isotonic predictions on their \CalibYear{} fitting population and are not independent calibration tests. All rows use non-held-out validation cells. Higher AP is better; lower Brier and ECE are better.}\label{tab:validation}\\
\toprule
Model & {Select AP} & {Calib AP} & {Calib Brier} & {Calib ECE} \\
\midrule\endfirsthead
\multicolumn{5}{l}{\tablename\ \ref{tab:validation} continued}\\
\toprule
Model & {Select AP} & {Calib AP} & {Calib Brier} & {Calib ECE} \\
\midrule\endhead
\bottomrule\endfoot
\multicolumn{5}{@{}l}{\textit{Full strict}}\\*
Climatology & \StrictClimatologySelectionAP & \StrictClimatologyCalibrationAP & \StrictClimatologyCalibrationBrier & \StrictClimatologyCalibrationECE \\*
Persistence & \StrictPersistenceSelectionAP & \StrictPersistenceCalibrationAP & \StrictPersistenceCalibrationBrier & \StrictPersistenceCalibrationECE \\*
Logistic regression & \StrictLogisticSelectionAP & \StrictLogisticCalibrationAP & \StrictLogisticCalibrationBrier & \StrictLogisticCalibrationECE \\*
Selected LightGBM & \StrictBoostSelectionAP & \StrictBoostCalibrationAP & \StrictBoostCalibrationBrier & \StrictBoostCalibrationECE \\[4pt]
\multicolumn{5}{@{}l}{\textit{Geographic features}}\\*
Climatology & \GeoClimatologySelectionAP & \GeoClimatologyCalibrationAP & \GeoClimatologyCalibrationBrier & \GeoClimatologyCalibrationECE \\*
Persistence & \GeoPersistenceSelectionAP & \GeoPersistenceCalibrationAP & \GeoPersistenceCalibrationBrier & \GeoPersistenceCalibrationECE \\*
Logistic regression & \GeoLogisticSelectionAP & \GeoLogisticCalibrationAP & \GeoLogisticCalibrationBrier & \GeoLogisticCalibrationECE \\*
Selected LightGBM & \GeoBoostSelectionAP & \GeoBoostCalibrationAP & \GeoBoostCalibrationBrier & \GeoBoostCalibrationECE \\[4pt]
\end{longtable}

\endgroup

\FloatBarrier
\TableHeadingSpace
\subsection{Annual and Santa Cruz scores}\label{app:subset-scores}

\begingroup\footnotesize
\setlength{\tabcolsep}{3pt}
\begin{longtable}{@{}lS[table-format=1.3]S[table-format=1.3]S[table-format=1.4]S[table-format=1.3]S[table-format=1.3]@{}}
\caption{Temporal-model breakdowns by issue year and for Santa Cruz, restricted to eligible non-held-out cells. Each group gives its shared evaluated row count. Annual and regional intervals are not recorded. Regional budgets use Santa Cruz's eligible cell area.}\label{tab:subsets}\\
\toprule
Model & {AP} & {Brier} & {ECE} & {R@\SmallBudget{}\%} & {R@\LargeBudget{}\%} \\
\midrule\endfirsthead
\multicolumn{6}{l}{\tablename\ \ref{tab:subsets} continued}\\
\toprule
Model & {AP} & {Brier} & {ECE} & {R@\SmallBudget{}\%} & {R@\LargeBudget{}\%} \\
\midrule\endhead
\bottomrule\endfoot
\multicolumn{6}{@{}l}{\textit{\YearA{} (\YearAClimatologyRows{} rows)}}\\*
Climatology & \YearAClimatologyAP & \YearAClimatologyBrier & \YearAClimatologyECE & \YearAClimatologyRecallSmall & \YearAClimatologyRecallLarge \\*
Persistence & \YearAPersistenceAP & \YearAPersistenceBrier & \YearAPersistenceECE & \YearAPersistenceRecallSmall & \YearAPersistenceRecallLarge \\*
Logistic regression & \YearALogisticAP & \YearALogisticBrier & \YearALogisticECE & \YearALogisticRecallSmall & \YearALogisticRecallLarge \\*
LightGBM & \YearABoostAP & \YearABoostBrier & \YearABoostECE & \YearABoostRecallSmall & \YearABoostRecallLarge \\[4pt]
\multicolumn{6}{@{}l}{\textit{\YearB{} (\YearBClimatologyRows{} rows)}}\\*
Climatology & \YearBClimatologyAP & \YearBClimatologyBrier & \YearBClimatologyECE & \YearBClimatologyRecallSmall & \YearBClimatologyRecallLarge \\*
Persistence & \YearBPersistenceAP & \YearBPersistenceBrier & \YearBPersistenceECE & \YearBPersistenceRecallSmall & \YearBPersistenceRecallLarge \\*
Logistic regression & \YearBLogisticAP & \YearBLogisticBrier & \YearBLogisticECE & \YearBLogisticRecallSmall & \YearBLogisticRecallLarge \\*
LightGBM & \YearBBoostAP & \YearBBoostBrier & \YearBBoostECE & \YearBBoostRecallSmall & \YearBBoostRecallLarge \\[4pt]
\multicolumn{6}{@{}l}{\textit{\YearC{} (\YearCClimatologyRows{} rows)}}\\*
Climatology & \YearCClimatologyAP & \YearCClimatologyBrier & \YearCClimatologyECE & \YearCClimatologyRecallSmall & \YearCClimatologyRecallLarge \\*
Persistence & \YearCPersistenceAP & \YearCPersistenceBrier & \YearCPersistenceECE & \YearCPersistenceRecallSmall & \YearCPersistenceRecallLarge \\*
Logistic regression & \YearCLogisticAP & \YearCLogisticBrier & \YearCLogisticECE & \YearCLogisticRecallSmall & \YearCLogisticRecallLarge \\*
LightGBM & \YearCBoostAP & \YearCBoostBrier & \YearCBoostECE & \YearCBoostRecallSmall & \YearCBoostRecallLarge \\[4pt]
\multicolumn{6}{@{}l}{\textit{Santa Cruz (\SantaCruzClimatologyRows{} rows)}}\\*
Climatology & \SantaCruzClimatologyAP & \SantaCruzClimatologyBrier & \SantaCruzClimatologyECE & \SantaCruzClimatologyRecallSmall & \SantaCruzClimatologyRecallLarge \\*
Persistence & \SantaCruzPersistenceAP & \SantaCruzPersistenceBrier & \SantaCruzPersistenceECE & \SantaCruzPersistenceRecallSmall & \SantaCruzPersistenceRecallLarge \\*
Logistic regression & \SantaCruzLogisticAP & \SantaCruzLogisticBrier & \SantaCruzLogisticECE & \SantaCruzLogisticRecallSmall & \SantaCruzLogisticRecallLarge \\*
LightGBM & \SantaCruzBoostAP & \SantaCruzBoostBrier & \SantaCruzBoostECE & \SantaCruzBoostRecallSmall & \SantaCruzBoostRecallLarge \\[4pt]
\end{longtable}

\endgroup

\FloatBarrier
\RainfallTableSpace
\subsection{Exploratory secondary rainfall intervals}\label{app:rainfall-secondary}
Lower Brier and ECE favor rainfall; lower ROC-AUC favors strict. These comparisons have no multiplicity adjustment and do not change the AP-based decisions.

\begin{table}[htbp]
\centering\small
\caption{Saved exploratory rainfall-minus-strict secondary intervals from common month-block resamples, conditional on fitted models and shared finite-prediction support.}\label{tab:rainfall-secondary}
\begin{tabular}{@{}lc@{}S[table-format=-1.7]@{,\,}S[table-format=-1.7]@{}c@{}}
\toprule
Metric & \multicolumn{4}{c}{\ConfidencePercent{}\% difference interval} \\
\midrule
Brier score (lower is better) & $[$ & \RainfallBrierLower & \RainfallBrierUpper & $]$ \\
ECE (lower is better) & $[$ & \RainfallECELower & \RainfallECEUpper & $]$ \\
ROC-AUC (higher is better) & $[$ & \RainfallAUCLower & \RainfallAUCUpper & $]$ \\
\bottomrule
\end{tabular}

\end{table}

\FloatBarrier
\subsection{Reproduction and experimental deployment}\label{app:reproduction}
\begin{samepage}
The research environment and commands for reproducing results are:
\begin{quote}\begin{minipage}{\linewidth}\small\setlength{\topsep}{0pt}
\begin{verbatim}
uv sync --all-extras
uv run --no-sync bolivia-fire acquire boundaries
uv run --no-sync bolivia-fire build cells
uv run --no-sync bolivia-fire build detections
uv run --no-sync bolivia-fire build panel
uv run --no-sync bolivia-fire train --variant strict
uv run --no-sync bolivia-fire select --variant strict
\end{verbatim}
\end{minipage}\end{quote}
\end{samepage}
The archived panel and covariates support direct reuse. Recalculation from saved predictions is limited to the strict primary selection, calibration and test exports and the alternative-learner vectors. Rainfall, sensitivity and environmental-covariate results can be verified from archived result records and fitted models; regenerating their test predictions requires the logged test-access procedure. Recomputing any final-test metric is a logged test opening governed by Section~\ref{sec:test-guard}. Validation-only evaluation uses \verb|--split val_select| or \verb|--split val_calib|; final-test evaluation additionally requires \verb|--open-final-test|.

Rebuilding from raw inputs is partially supported. Restore the archived snapshots into their repository-relative paths, including the checksum-identified outage HTML, and retrieve the provider-only inputs identified in the availability statement. MapBiomas re-extraction requires an Earth Engine account; FIRMS Area API access requires a local credential. Fresh downloads can differ from the recorded inputs. Full panel rebuilding omits issue-date bounds because a restricted rebuild replaces the requested year's output. Follow the acquisition and covariate documentation for rainfall and environmental features; missing covariates stop fitting rather than silently reducing the input population. Retain the recorded package and accelerator versions.

\paragraph{Experimental deployment.} \url{https://bolivia-fire.cupesi.org} provides experimental weekly near-real-time (NRT) estimates, immutable forecast records and a verification workflow on a separate live-data branch. It applies the strict temporal model across the grid; the separately selected geographic experiment does not validate that deployment. Initialized and backfilled records are excluded from on-time prospective summaries. Near-real-time detections do not contain the source-type field used to distinguish presumed vegetation fires in the standard-processing data. The live system approximates the vegetation filter with a \SupplementMaskMetres{}\,m buffer around known non-vegetation/static standard-processing detections through the recorded cutoff; this can remove nearby vegetation detections or miss new static sources. The research software documentation in \path{docs/live.md} records that cutoff, input provenance and prospective/backfill rules. Province, municipality and locality labels provide display context only. Historical scores do not establish operational warning utility.

\FloatBarrier
\TableHeadingSpace
\subsection{Sequential exclusions}\label{app:exclusions}
Table~\ref{tab:exclusions} reports the recorded fitting and test exclusions. The geographic-membership exclusion removes held-out cells for ordinary partitions and non-held-out cells for the geographic test.

\begin{table}[htbp]
\centering\footnotesize
\setlength{\tabcolsep}{3pt}
\caption{Recorded sequential exclusions. Scanned includes all globally purged rows as well as the requested partition; Purged therefore repeats that global count, unlike the within-period descriptive partition totals. Remaining exclusions proceed through missing label, other geographic population, incomplete recent observation history and the optional low-coverage filter. Eligible rows receive metrics or fitting. The optional low-coverage filter excluded no rows in these populations.}\label{tab:exclusions}
\begin{tabular}{@{}lS[table-format=8.0]S[table-format=6.0]S[table-format=6.0]S[table-format=7.0]S[table-format=7.0]S[table-format=8.0]@{}}
\toprule
Population & {Scanned} & {Purged} & {Missing} & {Other geo.} & {History} & {Eligible} \\
\midrule
Train & \ExclusionTrainScanned & \ExclusionTrainPurged & \ExclusionTrainMissing & \ExclusionTrainGeography & \ExclusionTrainHistory & \ExclusionTrainEligible \\
Selection & \ExclusionSelectionScanned & \ExclusionSelectionPurged & \ExclusionSelectionMissing & \ExclusionSelectionGeography & \ExclusionSelectionHistory & \ExclusionSelectionEligible \\
Calibration & \ExclusionCalibrationScanned & \ExclusionCalibrationPurged & \ExclusionCalibrationMissing & \ExclusionCalibrationGeography & \ExclusionCalibrationHistory & \ExclusionCalibrationEligible \\
Temporal test & \ExclusionTemporalScanned & \ExclusionTemporalPurged & \ExclusionTemporalMissing & \ExclusionTemporalGeography & \ExclusionTemporalHistory & \ExclusionTemporalEligible \\
Geographic test & \ExclusionGeographicScanned & \ExclusionGeographicPurged & \ExclusionGeographicMissing & \ExclusionGeographicGeography & \ExclusionGeographicHistory & \ExclusionGeographicEligible \\
\bottomrule
\end{tabular}

\end{table}

\FloatBarrier
\subsection{Vegetation and land-cover processing}\label{app:covariate-processing}
AppEEARS inputs retain native sinusoidal pixels and source-period, production and checksum metadata. NDVI/EVI scaling is verified before conversion; quality and observation dates remain unscaled. Both indices share reliability, MODLAND, usefulness, aerosol, land and cloud/snow/shadow screening. Accepted pixel means use WGS84 geodesic pixel-cell-national intersection areas, while valid-area fraction uses the complete national cell area and must reach \ExtensionValidArea{}.

Composites become eligible at midnight following their inclusive final day plus \ExtensionLag{} days; a separately fitted \ExtensionLongLag{}-day lag is a sensitivity. Current and change-reference observations must be at most \ExtensionCurrentAge{} and \ExtensionReferenceAge{} days old, respectively. Predictors include means, changes, training-season anomalies, valid fraction and age. Anomalies use at least \ExtensionReferenceYears{} training reference years and leave-one-year-out training references; geographic references pool non-held-out departmental cells.

MapBiomas class masks are averaged within project cells on a latitude--longitude grid and normalized over mapped classes. This approximates geodesic pixel and national-boundary intersections; it does not compute them exactly. The latest accessible map preceding the issue year is used only if it is at most \ExtensionLandAge{} years old. In an independent \num{\LandPilotCells}-cell polygon-intersection pilot, the maximum class-fraction and mapped-coverage differences were \LandPilotFractionError{} and \LandPilotCoverageError{}, respectively, below the \LandPilotTolerance{} acceptance tolerance; these pilot results are not a national error bound. Nodata, unobserved and unmapped codes are excluded; mapped coverage must reach \ExtensionLandCoverage{}. Forest, nonforest vegetation, farming, nonvegetated and water fractions sum to one over mapped area. A preceding map year does not undo smoothing that uses later years. The processing rules and exclusion records are specified in \path{docs/vegetation-acquisition.md}, \path{docs/covariates.md} and the corresponding saved evaluation summaries. The vegetation selection is in \path{artifacts/evaluation/extension_selection.json}; amended land-cover selections and exploratory test summaries are in the \path{artifacts/evaluation/amendments/} subdirectories.

\CovariateTableSpace
\subsection{Secondary and exploratory covariate comparisons}\label{app:covariate-secondary}
\begin{table}[htbp]
\centering\footnotesize
\setlength{\tabcolsep}{4pt}
\caption{Secondary covariate evidence and exploratory temporal-test intervals. Geographic-feature validation uses non-held-out cells. Intervals are \ConfidencePercent{}\% calendar-month-block intervals; decisions use unrounded endpoints, and signed zero retains the direction of a nonzero endpoint. These rows do not change the primary decisions in Table~\ref{tab:environmental-covariates}.}\label{tab:covariate-secondary}
\begin{tabular}{@{}lrS[table-format=7.0]cl@{}}
\toprule
Contrast & Lag (d) & {Paired rows} & 95\% CI for AP difference & Interpretation \\
\midrule
\multicolumn{5}{@{}l}{\textit{Temporal lag sensitivity}}\\[2pt]
H3 & \ExtensionRunBLag{} & \ExtensionDiagnosticBRows & $[$\(\hphantom{-}\ExtensionDiagnosticBLow\),\,\(\ExtensionDiagnosticBHigh\)$]$ & Positive secondary interval \\
H5 & \ExtensionRunELag{} & \ExtensionDiagnosticERows & $[$\(\ExtensionDiagnosticELow\),\,\(\ExtensionDiagnosticEHigh\)$]$ & Inconclusive sensitivity \\
\addlinespace[4pt]
\multicolumn{5}{@{}l}{\textit{Geographic-feature validation}}\\[2pt]
H3 & \ExtensionRunFLag{} & \ExtensionDiagnosticFRows & $[$\(\ExtensionDiagnosticFLow\),\,\(\ExtensionDiagnosticFHigh\)$]$ & Inconclusive secondary interval \\
H3 & \ExtensionRunGLag{} & \ExtensionDiagnosticGRows & $[$\(\ExtensionDiagnosticGLow\),\,\(\ExtensionDiagnosticGHigh\)$]$ & Inconclusive secondary interval \\
H4 & n/a & \ExtensionDiagnosticHRows & $[$\(\hphantom{-}\ExtensionDiagnosticHLow\),\,\(\ExtensionDiagnosticHHigh\)$]$ & Positive secondary interval \\
H5 & \ExtensionRunILag{} & \ExtensionDiagnosticIRows & $[$\(\ExtensionDiagnosticILow\),\,\(\ExtensionDiagnosticIHigh\)$]$ & Inconclusive secondary interval \\
H5 & \ExtensionRunJLag{} & \ExtensionDiagnosticJRows & $[$\(\hphantom{-}\ExtensionDiagnosticJLow\),\,\(\ExtensionDiagnosticJHigh\)$]$ & Positive secondary interval \\
\addlinespace[4pt]
\multicolumn{5}{@{}l}{\textit{Exploratory temporal test}}\\[2pt]
H3 & \ExtensionLag{} & \ExtensionTestARows & $[$\(\ExtensionTestALow\),\,\(\ExtensionTestAHigh\)$]$ & Exploratory \\
H4 & n/a & \ExtensionTestBRows & $[$\(\hphantom{-}\ExtensionTestBLow\),\,\(\ExtensionTestBHigh\)$]$ & Exploratory \\
H5 & \ExtensionLag{} & \ExtensionTestCRows & $[$\(\hphantom{-}\ExtensionTestCLow\),\,\(\ExtensionTestCHigh\)$]$ & Exploratory \\
\bottomrule
\end{tabular}

\end{table}

\end{document}